\documentclass[preprint]{elsarticle}

\usepackage{lineno}
\modulolinenumbers[5]

\usepackage{slashbox}
\usepackage{graphicx}
\usepackage{dcolumn}
\usepackage{bm}
\usepackage{references}
\usepackage{hyperref} 

\usepackage{equationarray}
\usepackage[normalem]{ulem}
\usepackage[usenames]{xcolor}
\usepackage{booktabs}

\newcommand{\be}{\begin{equation}}
\newcommand{\ee}{\end{equation}}
\newcommand{\ba}{\begin{eqnarray}}
\newcommand{\ea}{\end{eqnarray}}
\newcommand{\bd}{\begin{displaymath}}
\newcommand{\ed}{\end{displaymath}}

\newcommand{\mm}{\mathcal{M}}

\usepackage{ulem} 

\journal{Annals of Physics}

\begin{document}

\begin{frontmatter}

\title{Degenerate limit thermodynamics beyond leading order for models of dense matter}

\author{Constantinos Constantinou\fnref{fncc}}
\address{Institute for Advanced Simulation, Institut f\"{u}r Kernphysik, and J\"{u}lich Center \\
for Hadron Physics, Forschungszentrum J\"{u}lich, D-52425 J\"{u}lich, Germany}
\fntext[fncc]{c.constantinou@fz-juelich.de }

\author{Brian Muccioli\fnref{fnbm}}
\address{Department of Physics and Astronomy, Ohio University, Athens, OH 45701}
\fntext[fnbm]{bm956810@ohio.edu}

\author{Madappa Prakash\fnref{fnmp}}
\address{Department of Physics and Astronomy, Ohio University, Athens, OH 45701}
\fntext[fnmp]{prakash@ohio.edu}

\author{James M. Lattimer\fnref{fnjl}}
\address{Department of Physics and Astronomy, Stony Brook University, Stony Brook, NY 11794-3800}
\fntext[fnjl]{james.lattimer@stonybrook.edu}

\begin{abstract}

Analytical formulas for next-to-leading order temperature corrections
to the thermal state variables of interacting nucleons in bulk matter are derived in
the degenerate limit. The formalism developed is applicable to a wide
class of non-relativistic and relativistic models of hot and dense
matter currently used in nuclear physics and
astrophysics (supernovae, proto-neutron stars and neutron star
mergers) as well as in condensed matter physics.  We consider the
general case of arbitrary dimensionality of momentum space and an
arbitrary degree of relativity (for relativistic mean-field theoretical
models).  
For non-relativistic zero-range
interactions, knowledge of 
the Landau
effective mass  suffices to compute next-to-leading order
effects, but in the case of finite-range interactions, momentum
derivatives of the Landau effective mass function
up to second order are required.  
Numerical computations are
performed to compare results from our analytical formulas with the exact
results for zero- and finite-range potential and relativistic
mean-field theoretical models.  In all cases, inclusion of
next-to-leading order temperature effects 
substantially extends the
ranges of partial degeneracy for which the analytical treatment remains
valid.

\end{abstract}



\end{frontmatter}

\linenumbers


\section{INTRODUCTION}
\label{Sec:Intro}

Homogeneous bulk matter comprised of fermions is commonly encountered
in astrophysics, condensed matter physics, and nuclear physics.  For
extreme degenerate to near-degenerate conditions which prevail when
the temperature is small compared to the Fermi temperature, Landau's
Fermi Liquid Theory (FLT) has been a useful guide to describe the
thermodynamic and transport properties of the system (see, {\em e.g.,}
\cite{flt} and references therein).  The equation of state (EOS) of
dense matter in cold and catalyzed neutron stars, for example, is
dominated by the zero-temperature properties (which predominantly
determine the structure of and neutrino interactions within the star)
while finite-temperature corrections (important for the cooling of
neutron stars) are adequately given by the degenerate limit
expressions from FLT. The leading order FLT corrections to the energy
density and pressure are quadratic in the temperature; corrections to
the entropy and specific heats are linear in the temperature. However,
matter in supernovae and proto-neutron stars \cite{LB86,Burrows88},
especially in situations in which collapse to a
black hole occurs, may reach temperatures exceeding the Fermi
temperature, in which case the finite-temperature contributions extend
beyond those given by the FLT.  In 
neutron star mergers,
it is likely that in some cases a hyper-massive neutron star, or HMNS,
is formed: the merged remnant mass exceeds the cold maximum mass.  The
metastable support is provided by rotation, including differential
rotation, and thermal effects.  The timescale over which collapse to a
black hole eventually occurs, potentially observable in gravitational
wave signatures, will therefore be sensitive to thermal effects \cite{Sekiguchi11}.
In this contribution, we derive analytical formulas for
next-to-leading order temperature effects in the state variables of
interacting nucleons in both the non-relativistic and relativistic
limits for a variety of nuclear interaction models.

For non-relativistic models with zero-range interactions, knowledge of
the Landau effective mass is
sufficient to satisfy thermodynamic identities.  However, in the
general case of finite-range interactions, momentum derivatives of the
Landau effective mass function up to second order are required.  
We compare results from the analytical
expressions to exact numerical calculations for zero- and finite-range
potential models as well as for relativistic mean-field theoretical
models.  The analytic next-to-leading order expressions lead to an improvement of
the leading order results of FLT, as demonstrated by the wider ranges
of degeneracy and temperature for which they remain valid.  In
addition, we derive relations in a form that are independent of the
dimensionality of the momentum space under consideration.  Therefore,
although our discussion 
focuses on examples from dense
matter physics, which are three-dimensional systems in
momentum space, the expressions derived can have a wider application
to certain problems in condensed matter physics in which the momentum
space is two-dimensional.

The paper is organized as follows. In Sec. \ref{Sec:General}, the formalism to calculate next-to-leading order corrections to the 
results of FLT in D-dimensions is developed. Analytical formulas appropriate for 3-dimensions are given in Sec. \ref{Sec:D3}, whereas 
Sec. \ref{Sec:D2} contains results for 2-dimensions. The formalism is applied to zero- and finite-range potential models and a 
relativistic  field-theoretical model in Sec. \ref{Sec:Models}. Numerical results for these models are presented in Sec. \ref{Sec:Results} 
where the extent to which the next-to-leading order corrections improve the FLT results are demonstrated. Section \ref{Sec:Conclusions} 
presents  a summary and conclusions. Useful formulas for the evaluation of the thermal properties are provided in  Appendices A, B, and C.

\section{GENERAL CONSIDERATIONS}
\label{Sec:General}

For a generic Hamiltonian density $\mathcal{H}(n,\tau)$ where $n$ and $\tau$ are the number and kinetic energy densities respectively,  
the single-particle potential $U$
is obtained from a functional differentiation of $\mathcal{H}$ with respect to $n$, and can contain 
terms depending on $n$ as well as the momentum $p$: 
\be
U(n,p) = \frac{\delta \mathcal{H}}{\delta n} = \mathcal{U}(n) + R(p) \,,
\ee
where $ \mathcal{U}(n)$ denotes contributions that depend on $n$ only.
Note that $R$ above may also be $n$-dependent but we will suppress this for notational simplicity.

The study of the thermodynamic properties of a fermion system involves integrals of the form
\be
I = \int_0^{\infty}dp~ g(p) f(p)\,, \qquad f(p) = 
\left[1+\exp\left(\frac{\epsilon-\mu}{T}\right)\right]^{-1} \,,
\label{gofp}
\ee
where $T$ is the temperature, $\mu$ is the chemical potential, and the single-particle spectrum 
\be
\epsilon = \frac{p^2}{2m} + U(n,p) \,.
\ee
The structure of the function $g(p)$ is determined by the state variable under consideration.
In general, integrals involving the Fermi function $f(p)$
do not  admit  analytical solutions and thus require numerical treatment. In the low-temperature limit, however, 
when the degeneracy parameter
\be
\eta = \frac{\mu - \epsilon(p=0)}{T}
\ee
is large, these integrals can be approximately evaluated employing the Sommerfeld expansion (see, {\em e.g.,} \cite{LLI}) by transforming Eq. (\ref{gofp}) to 
\be
I = \int_0^{\infty}dy\frac{\phi(y)}{1+\exp(y-\eta)} \stackrel{\eta \gg 1}{\longrightarrow} \int_0^{\eta}\phi(y)~dy  
    + \frac{\pi^2}{6}\left.\frac{d\phi}{dy}\right|_{y=\eta} 
   + \frac{7\pi^4}{360}\left.\frac{d^3\phi}{dy^3}\right|_{y=\eta} + \ldots.   \label{sommerfeld}
\ee
with the identification $\phi(y)dy = g(p)dp$, and the substitution
\be
y \equiv \frac{\epsilon - \mathcal{U}(n)}{T} = \frac{p^2}{2mT} + \frac{R(p)}{T} \,,
\ee
from which it follows that
\ba
\frac{dy}{dp} = \frac{p}{\mm T}  \label{dydp}  \quad {\rm and} \quad
\phi(y) = \frac{\mm T}{p}g(p) \,, \label{fi} 
\ea
where the Landau effective mass function
\be
\mm(p) = m\left[1+\frac{m}{p}\frac{dR(p)}{dp}\right]^{-1} \,. \label{genm}
\ee
This function is implicitly temperature-dependent and its relation to the Landau effective mass $m^*$ is 
\be
\mm(p=p_F;T=0) = m^*\,,
\ee
where $p_F$ is the Fermi momentum. 
From the relations in Eq. (\ref{fi}) 
\raggedbottom
\ba
\frac{d\phi}{dy} &=& -\frac{T^2\mm^2g}{p^3}\left[1-p\left(\frac{g'}{g}+\frac{\mm'}{\mm}\right)\right] \label{dfidy}  
\raggedbottom
\ea
\raggedbottom
\ba
\frac{d^3\phi}{dy^3} &=& -\frac{15T^4\mm^4g}{p^7}\left[1-p\left(\frac{g'}{g}+\frac{5}{3}\frac{\mm'}{\mm}\right)\right. \nonumber \\
  &+& \frac{2}{5}p^2\left(\frac{g''}{g}+\frac{11}{3}\frac{\mm'g'}{\mm g}+\frac{11}{6}\frac{\mm'^2}{\mm^2}+\frac{7}{6}\frac{\mm''}{\mm}\right)
  \nonumber  \\
  &-&  \frac{p^3}{15}\left(\frac{g'''}{g}+7\frac{\mm'^2g'}{\mm^2g}+6\frac{\mm'g''}{\mm g}+4\frac{\mm''g}{\mm g}\right. \nonumber \\
  &+& \left.\left. \frac{\mm'^3}{\mm^3}+\frac{\mm'''}{\mm}+4\frac{\mm'\mm''}{\mm^2}\right)\right]  \,, 
    \label{d3fi}
\ea
\raggedbottom
where the primes denote differentiation with respect to $p$.

For a system in $D$ dimensions having $\gamma$ internal degrees of freedom, the number density is given by
\be
n = C_D \int dp~p^{D-1}~f_p \quad {\rm with} \quad C_D = \frac{\gamma}{(2\pi\hbar)^D}~\frac{D\pi^{D/2}}{(D/2)!} \label{numden}\,.
\ee
The combination of Eqs. (\ref{sommerfeld}),(\ref{dfidy}), and (\ref{d3fi}) with $g_n=p^{D-1}$ results in 
\ba
n &=& \frac{C_D}{D}\left\{p_{\mu}^D + \frac{\pi^2}{6}Dp_{\mu}^{D-4}\mm_{\mu}^2T^2\left(D-2+p_{\mu}\frac{\mm_{\mu}'}{\mm_{\mu}}\right)\right. \nonumber \\
  &+& \frac{7\pi^4}{360}Dp_{\mu}^{D-8}\mm_{\mu}^4T^4\left[(D-6)(D-4)(D-2)\frac{}{}\right.  \nonumber  \\
  &+&\left(\frac{p_{\mu}\mm_{\mu}'}{\mm_{\mu}}\right)^3 + (7D-18)\left(\frac{p_{\mu}\mm_{\mu}'}{\mm_{\mu}}\right)^2  \nonumber \\
  &+& (6D^2-40D+59)\frac{p_{\mu}\mm_{\mu}'}{\mm_{\mu}} + \frac{p_{\mu}^3\mm_{\mu}'''}{\mm_{\mu}}  \nonumber \\
  &+& \left.\left. 4~\frac{p_{\mu}^3\mm_{\mu}'\mm_{\mu}''}{\mm_{\mu}^2}+(4D-11)\frac{p_{\mu}^2\mm_{\mu}''}{\mm_{\mu}}\right] + \ldots \right\}   \,,
\label{seriesn}
\ea
where the subscript $\mu$ denotes quantities evaluated at $\epsilon=\mu$, {\it i.e.,}
\be
\epsilon = \mu = \frac{p_{\mu}^2}{2m} + R(p_{\mu}) + \mathcal{U}(n) \,.   \label{mu}
\ee
For $N$ particles in a volume $V$, the number density $n=N/V$ at $T=0$ and at finite $T$ is the same. 
Equating the result in Eq. (\ref{seriesn}) to its $T=0$ counterpart $n=C_Dp_F^3/D$, and perturbatively inverting  we get 
\be
p_{\mu} = p_F\left[1-\frac{\pi^2}{6}\frac{m^{*2}T^2}{p_F^4}\left(D-2+\frac{p_F\mm_F'}{m^*}\right)+\ldots \right] \label{pmu}
\ee
with
\ba
p_F = \left(\frac{nD}{C_D}\right)^{1/D}  \quad {\rm and} \quad 
\mm_F' = \left.\frac{d\mm}{dp}\right|_{p_F} \,.
\label{pF}
\ea

As our main goal here is to derive the next-to-leading order correction in temperature for the entropy density $s$, it suffices
to truncate the series expansion 
of $p_{\mu}$ to $\mathcal{O}(T^2)$.  We will show below that higher-order terms do not contribute at this level of 
approximation where we may also neglect the temperature dependence of $\mm$ and its derivatives. 
The result in Eq. (\ref{pmu}) helps us to work only with quantities defined on the Fermi surface as done in Landau's Fermi-Liquid theory 
\cite{flt,ll9,BaymChin76}.
The entropy density is formally given by
\be
s = -C_D \int dp~p^{D-1}[f_p\ln f_p - (1-f_p)\ln(1-f_p)] \,.
\label{sformal}
\ee
Integrating this expression twice by parts we obtain
\be
s = \frac{1}{T}\left\{\frac{\tau}{m}\left(\frac{1}{2}+\frac{1}{D}\right)+ n(\mathcal{U}-\mu)
  + C_D \int dp~p^{D-1}f_p\left[R(p)+\frac{p}{D}\frac{dR(p)}{dp}\right]\right\} 
  \label{sd1}
\ee
where 
\be
\tau = C_D \int dp~p^{D+1}f_p
\label{kinden}
\ee
is the kinetic energy density.
With the aid of Eq. (\ref{mu}) for the chemical potential, Eq. (\ref{sd1}) can be written as 
\be
s = \frac{1}{T}\left\{\frac{\tau}{m}\left(\frac{1}{2}+\frac{1}{D}\right)- n\frac{p_{\mu}^2}{2m}
  + C_D \int dp~p^{D-1}f_p\left[R(p)  - R(p_{\mu})+ \frac{p}{D}\frac{dR(p)}{dp}  \right]\right\} 
  \label{sd1a}  
\ee
from which we identify the functions
\ba
g_{1s} (p)&=& \frac{p^{D+1}}{m}\left(\frac{1}{2}+\frac{1}{D}\right)-p^{D-1}\frac{p_{\mu}^2}{2m}  \\ 
g_{2s} (p) &=& p^{D-1}\left[R(p)  - R(p_{\mu}) + \frac{p}{D}\frac{dR(p)}{dp}  \right]
\ea
to be used in the Sommerfeld expansion. For both of these functions, the first term on the right-hand side of Eq. (\ref{sommerfeld}) 
involving an integral vanishes yielding 
\ba
s &=& \frac{\pi^2}{3}C_Dp_{\mu}^{D-2}\mm_{\mu}T  
   + \frac{7\pi^4}{90}C_Dp_{\mu}^{D-6}\mm_{\mu}^3T^3\left[(D-4)(D-2) \frac{}{}\right.  \nonumber  \\
  &+& \left. p_{\mu}^2\frac{\mm_{\mu}'^2}{\mm_{\mu}^2}+p_{\mu}^2\frac{\mm_{\mu}''}{\mm_{\mu}}+(3D-7)p_{\mu}\frac{\mm_{\mu}'}{\mm_{\mu}}\right]  \,. 
  \label{sgen}
  \ea
Use of Eqs. (\ref{pmu}) and (\ref{pF}) in the above result delivers the working expression for $s$ in terms of quantities defined on the 
Fermi surface:   
\ba
s  &\simeq& \frac{\pi^2}{3}C_Dp_F^{D-2}m^*T 
  + \frac{\pi^4}{45}C_Dp_F^{D-6}m^{*3}T^3\left[(D-9)(D-2) \frac{}{}\right.  \nonumber  \\
  &+& \left. \frac{7}{2}p_F^2\frac{\mm_F'^2}{m^{*2}}+\frac{7}{2}p_F^2\frac{\mm_F''}{m^*}+\frac{(16D-39)}{2}p_F\frac{\mm_F'}{m^*}\right]\,,
\label{sd2}
\ea
where the ${\cal O}(T)$ term is the well known result from FLT. 
We note that a large number of cancellations
occur in obtaining Eqs. (\ref{sgen}) and (\ref{sd2}) despite the complexity of 
of Eqs. (\ref{dfidy}) and (\ref{d3fi}).
For a system composed of different kinds of particles the total entropy density is a sum of the contributions from the 
individual species where, in Eq. (\ref{sd2}), the Fermi momentum, the effective mass, and its derivatives all carry a particle-species 
index $i$.

Equation (\ref{sd2}) forms the basis from which other properties of the system can be derived. For example, the entropy per particle 
is the simple ratio $S=s/n$, whereas the thermal energy, pressure, and chemical potential can be obtained through the application of the appropriate Maxwell relations \cite{LLI}:
\ba
E_{th} = \int T~dS  \label{ed}\,, \quad 
P_{th} = -n^2 \int \frac{dS}{dn}dT \label{pd} \quad {\rm and} \quad 
\mu_{th} = -\int \frac{ds}{dn}dT  \label{mud}
\ea
[for a multiple-species system, $\mu_{th,i}=-\int (ds/dn_i)dT$].

The specific heats at constant volume and pressure are given by the standard thermodynamics expressions \cite{LLI}
\ba
C_V &=& T\left.\frac{\partial S}{\partial n}\right|_n = \left.\frac{\partial E_{th}}{\partial T}\right|_n    \label{cvd}\\
C_P &=& T\left.\frac{\partial S}{\partial n}\right|_P 
     = C_V + \frac{T}{n^2}\frac{\left(\left.\frac{\partial P_{th}}{\partial T}\right|_n\right)^2}{\left.\frac{\partial P}{\partial n}\right|_T}\,.
\label{cpd}
\ea

We note that the formalism above has not considered effects, for example,  from single particle-hole excitations, or from collective and pairing correlations near the Fermi surface \cite{flt,BF81,CB93}.  Contributions from these sources must be added to those considered here when appropriate.    

\section{RESULTS FOR D=3}
\label{Sec:D3}

For a single-species system of spin $1/2$ particles in 3 dimensions [for which $C_3=1/(\pi^2\hbar^3)$], the entropy density becomes
\be
s = \frac{p_Fm^*T}{3\hbar^3} - \frac{2\pi^2}{15\hbar^3}\frac{m^{*3}T^3}{p_F^3}(1-L_F) \,, \label{s31}
\ee
where 
\be
L_F \equiv \frac{7}{12}p_F^2\frac{\mm_F'^2}{m^{*2}} + \frac{7}{12}p_F^2\frac{\mm_F''}{m^*} + \frac{3}{4}p_F\frac{\mm_F'}{m^*} \,.
\label{LF}
\ee
We stress that, in general,
\be
\left.\frac{d\mm(p)}{dp}\right|_{p_F} = \mm_F' \ne m^{*\prime}=\frac{d\mm(p_F)}{dp_F}
\ee
as $R$ can contain both $p$ and $p_F$ (via $n$).
In terms of the level-density parameter $a=\pi^2m^*/(2p_F^2)=\pi^2/(4T_F)$ (where $T_F$ is the Fermi temperature), Eq. (\ref{s31}) can be written as 
\be
s = 2anT - \frac{16}{5\pi^2}a^3nT^3(1-L_F) \label{s32}.
\ee
The quantity $L_F$ arises from nontrivial momentum dependencies in the single-particle potential. 
For free gases (where $R(p)=0$), and for systems having 
only contact interactions where $R(p) \propto p^2$ (such as Skyrme models), $L_F=0$. 

Equation (\ref{s32}) in conjunction with Eqs. (\ref{mud})-(\ref{cpd}) leads to
\ba
S &=& 2aT -\frac{16}{5\pi^2}a^3T^3(1-L_F)\,, \quad E_{th} = aT^2 - \frac{12}{5\pi^2}a^3T^4(1-L_F)  
\label{sa3}\\
E_{th} &=& aT^2 - \frac{12}{5\pi^2}a^3T^4(1-L_F)  \\
P_{th} &=& \frac{2}{3}anQT^2    
      - \frac{8}{5\pi^2}a^3nQT^4\left(1-L_F+\frac{n}{2Q}\frac{dL_F}{dn}\right)  \label{pth3} \\
\mu_{th} &=& -a\left(1-\frac{2Q}{3}\right)T^2 
        + \frac{4}{5\pi^2}a^3T^4\left[(1-L_F)(1-2Q)-n\frac{dL_F}{dn}\right] \\
C_V &=& 2aT -\frac{48}{5\pi^2}a^3T^3(1-L_F)   \quad {\rm and} \quad
C_P = C_V + \frac{16}{9}\frac{a^2Q^2T^3}{\frac{dP_0}{dn}}  \,,  \label{cp3}
\ea
where
\be
Q = 1-\frac{3n}{2m^*}\frac{dm^*}{dn} \,.
\ee
In the derivation of Eq. (\ref{cp3}) we have assumed that the zero-temperature pressure $P_0$ is such that 
$dP_0/dn \gg dP_{th}/dn$. This condition will not be met in situations where $P_0$ is relatively flat  as in the 
vicinity of a critical point. When this is the case, we must use Eq. (\ref{cpd}) for $C_P$, with 
\ba
\left(\left.\frac{\partial P_{th}}{\partial T}\right|_n\right)^2 &=&
   \left(\frac{4}{3}anQT\right)^2 
   \left[1 - \frac{48}{5\pi^2}a^2T^2\left(1-L_F+\frac{n}{2Q}\frac{dL_F}{dn}\right)\right]  \\
\left.\frac{\partial P}{\partial n}\right|_T &=& 
     \frac{dP_0}{dn}+ \frac{2}{3}aQT^2\left(1-\frac{2Q}{3}+n\frac{dQ}{dn}\right)  \nonumber \\
    &-& \frac{5}{8\pi^2}a^3QT^4\left[\left(1-2Q+n\frac{dQ}{dn}\right)
     \left(1-L_F+\frac{n}{2Q}\frac{dL_F}{dn}\right)\right.   \nonumber \\
       &-& \left. n\frac{dL_F}{dn}\left(1-2Q+\frac{n}{2Q^2}\frac{dQ}{dn}\right)+\frac{n^2}{2Q}\frac{d^2L_F}{dn^2}\right] \,.
     \label{dpdn3} 
\ea
Similar considerations as with Eq. (\ref{cp3}) hold for the ratio of the specific heats
\be
\frac{C_P}{C_V} = 1+\frac{8}{9}\frac{aQ^2T^2}{\frac{dP_0}{dn}}  \label{cratio} \,.
\ee
Other quantities of interest in astrophysical applications include the thermal index $\Gamma_{th}$
\be
\Gamma_{th} = 1+\frac{P_{th}}{nE_{th}} = 1+\frac{2Q}{3}-\frac{4}{5\pi^2}a^2nT^2\frac{dL_F}{dn}
\ee
and the adiabatic index $\Gamma_S$
\be
\Gamma_S = \frac{C_P}{C_V}\frac{n}{P}\left.\frac{\partial P}{\partial n}\right|_T   
 = \frac{n}{P_0}\left[\frac{dP_0}{dn}+\frac{2}{3}aQT^2\left(1+\frac{2Q}{3}+n\frac{dQ}{dn}-\frac{n}{P_0}\frac{dP_0}{dn}\right)\right]
\ee
where, in addition to Eqs. (\ref{pth3}),(\ref{dpdn3}) and (\ref{cratio}), the approximation
\be
\frac{1}{P} \simeq \frac{1}{P_0}\left(1-\frac{P_{th}}{P_0}\right)  \label{pappr}
\ee
was used. In its native variables $(n,S)$, $\Gamma_S$ is given by
\be
\Gamma_S = \frac{n}{P}\left.\frac{\partial P}{\partial n}\right|_S  
 = \frac{n}{P_0+\frac{nQS^2}{6a}}\left[\frac{dP_0}{dn}+\frac{QS^2}{6a}\left(1+\frac{2}{3}Q+\frac{n}{Q}\frac{dQ}{dn}\right)\right].
\label{gammas1}  
\ee
To arrive to Eq. (\ref{gammas1}) one begins by inverting Eq. (\ref{sa3}) for the small parameter 
\be
aT = \frac{S}{2}+\frac{S^3}{5\pi^2}(1-L_F)
\ee
which is then employed in the expression for the thermal pressure with the results
\ba
P_{th} &=& \frac{nQ}{6a}S^2 + \frac{nQ}{30\pi^2a}S^4\left(1-L_F-\frac{3n}{2Q}\frac{dL_F}{dn}\right)   \\
\left.\frac{\partial P_{th}}{\partial n}\right|_S &=& \frac{Q}{6a}S^2\left(1+\frac{2}{3}Q+\frac{n}{Q}\frac{dQ}{dn}\right)  \nonumber \\
  &+& \frac{Q}{30\pi^2a}S^4\left[\left(1+\frac{2}{3}Q+\frac{n}{Q}\frac{dQ}{dn}\right)(1-L_F)\right. \nonumber \\
  &-& \left. 2n\frac{dL_F}{dn}\left(1+\frac{3}{2Q}\right)-\frac{3n^2}{2Q}\frac{d^2L_F}{dn^2}\right].
\ea
Finally, the result is truncated to $\mathcal{O}(S^2)$ in both the numerator as well as the denominator.
We refrain from invoking approximation (\ref{pappr}) 
as for nuclear systems, $P_0$ can cross 0 at low densities. This is
not a problem in the variables $(n,T)$ because the degenerate approximation breaks down at sufficiently low density regardless
of $T$. In the variables $(n,S)$, however, for small values of the entropy the system remains degenerate irrespective
of the density, and thus division by zero is avoided (as could happen if Eq. (\ref{pappr}) is used).  

We point out that the adiabatic index is related to the squared speed of sound $c_s$ according to
\be
\left(\frac{c_s}{c}\right)^2 = \Gamma_S \frac{P}{h+mn} \,,
\ee
where $h=nE+P$ is the enthalpy density.

\section{RESULTS FOR D=2}
\label{Sec:D2}

In condensed matter physics, 2-dimensional systems are of much interest.  
In the current framework, the entropy density is 
\be
s = \frac{\pi^2}{3}C_2m^*T + \frac{7\pi^4}{90}C_2\frac{m^{*3}T^3}{p_F^4} 
   \left(p_F^2\frac{\mm_F'^2}{m^{*2}} + p_F^2\frac{\mm_F''}{m^*} - p_F\frac{\mm_F'}{m^*} \right) \,
   \label{sD2}
\ee
with $C_2=(1/2\pi\hbar^2)$. 
A noteworthy feature of this result is that the  $\mathcal{O}(T^3)$ term receives contributions only from the derivatives 
of the effective mass function with respect to $p$ at the Fermi surface. Thus, it is absent not only for free gases but also 
for systems with contact interactions where the $p^2$-dependence of $R$ implies that $d\mm/dp = 0$. \\
In terms of the level density parameter $a=\pi^2m^*/(2p_F^2)$, 
and 
\ba
p_F &=& \left(\frac{2n}{C_2}\right)^{1/2}\,,  \qquad Q = 1 - \frac{n}{m^*}\frac{dm^*}{dn} \\
L_F &=& p_F^2\frac{\mm_F^{\prime 2}}{m^{*2}} + p_F^2\frac{\mm_F^{\prime\prime}}{m^*} - p_F\frac{\mm_F^{\prime}}{m^*}  
\ea
Eq. (\ref{sD2}) leads to 
\ba
S &=& \frac{4}{3}aT + \frac{56}{45\pi^2}a^3T^3L_F \,,  \qquad
E_{th} = \frac{2}{3}aT^2 + \frac{14}{15\pi^2}a^3T^4L_F  \\
P_{th} &=& \frac{2}{3}anQT^2 + \frac{14}{15\pi^2}a^3nQT^4\left(L_F-\frac{n}{3Q}\frac{dL_F}{dn}\right)  \\
\mu_{th} &=& -\frac{2}{3}aT^2(1-Q) - \frac{14}{45\pi^2}a^3T^4\left[L_F(1-3Q)+n\frac{dL_F}{dn}\right]  \\
C_V &=& \frac{4}{3}aT + \frac{56}{15\pi^2}a^3T^3L_F   \qquad {\rm and} \qquad
C_P = C_V + \frac{16}{9}\frac{a^2Q^2T^3}{\frac{dP_0}{dn}} \,.
\ea
The above results do not include the effects of collective excitations near the Fermi surface or of non-analytic contributions. As pointed out in Ref. \cite{CB93}, 2-dimensional Fermi systems in condensed matter physics (even with contact interactions) have $T^2$ contributions to the entropy from interactions separate
from those due to the collective modes. These $T^2$ contributions arise from non-analytic corrections to the real part of the self-energy.

\section{APPLICATION TO MODELS}
\label{Sec:Models}

In what follows, we compare the analytical results from the leading order corrections to Landau Fermi-liquid theory 
to the results of exact numerical calculations of the thermal state variables. 
These comparisons are made using models that are widely used in nuclear and neutron star phenomenology.
In the category of non-relativistic potential models, we begin with the
model, referred to as MDI(A), that reproduces the empirical properties of isospin symmetric
and asymmetric bulk nuclear matter \cite{cons15}, optical model fits to
nucleon-nucleus scattering data \cite{Hama:90}, heavy-ion flow data in the energy
range 0.5-2 GeV/A \cite{Danielewicz02}, and the largest well-measured neutron star mass of
2 $\rm{M}_\odot$ \cite{Demorest,Antoniadis}.  This model, which is based on Refs. \cite{Welke88,Das03}, incorporates finite range
interactions through a Yukawa-type, finite-range force, is contrasted
with a conventional zero-range Skyrme model known as SkO$^\prime$ \cite{Reinhard99}.  Both models predict nearly
identical zero-temperature properties at all densities and proton
fractions, including the neutron star maximum mass, but differ in their
predictions for heavy-ion flow data \cite{Prakash88b}. 
To provide a contrast, we also investigate a relativistic mean-field theoretical (MFT) model \cite{cons15} which yields zero-temperature 
properties similar to those of the two non-relativistic models chosen here. For all three models, we consider nucleonic 
matter in its pure neutron-matter (PNM, with $\gamma=2$) and symmetric nuclear matter (SNM, with $\gamma=4$)  configurations.

\subsection{Finite-range potential models}

 For the MDI(A) model \cite{Das03,cons15},  
the momentum-dependent part of the single-particle potential is given by
\ba
R(p) &=& \frac{2C_{\gamma}}{n_0}\frac{2}{(2\pi\hbar)^3}\int d^3p'~f_{p'}\frac{1}{1+\left(\frac{\vec p - \vec p'}{\Lambda}\right)^2}   \\
 &\stackrel{T=0}{\longrightarrow}& \frac{C_{\gamma}}{n_0}\frac{\Lambda^3}{\pi^2\hbar^3}\left\{\frac{p_F}{\Lambda} 
    - \arctan\left(\frac{p+p_F}{\Lambda}\right)+\arctan\left(\frac{p-p_F}{\Lambda}\right)\right.    \nonumber \\
    &+& \left.\frac{(\Lambda^2+p_F^2-p^2)}{4\Lambda p}\ln\left[\frac{\Lambda^2+(p+p_F)^2}{\Lambda^2+(p-p_F)^2}\right]\right\}.
\ea
For the coefficients $n_0$, $C_2$, $C_4$ and $\Lambda$ we use the values 0.16 fm$^{-3}$, -23.06 MeV, -128.9 MeV and 420.9 MeV, respectively.
Explicit expressions for the derivatives of $R(p)$ and their connection with $\mm$ and $L_F$ are provided in Appendix A. The
MDI Hamiltonian density is shown in Appendix B. For details of the exact numerical calculations, see Ref. \cite{cons15}.

\subsection{Zero-range Skyrme models}

Zero-range Skyrme models belong to that subset of the $D=3$ case for which $L_F=0$. This is because, for these models, the
momentum-dependent part of the potential has the form
\be
R = \beta(n)p^2
\ee
($\beta(n)$ is a density dependent factor) which renders the generalized effective mass to be $p-$independent:
\be
\mm = \frac{m}{1+\frac{m}{p}\frac{dR}{dp}} = \frac{1}{1+2m\beta(n)} \,,
\ee
and therefore its derivatives $\mm'=\mm''=0$. Hence $L_F=0$ as well. 
Consequently, the results in Sec. \ref{Sec:D3} for Skyrme models simplify considerably.
Results to be shown here are for the SKO$^\prime$ model \cite{Reinhard99}, the exact numerical calculations for which are described in Ref. \cite{cons15}.

\subsection{Relativistic models}
\label{Sec:MFT}

The single-particle energy spectrum of relativistic mean-field theoretical models \cite{Muller96}
obtained from the nucleon equation of motion has the structure
\be
\epsilon = E^* + U(n)\,, \quad E^* = [p^2 + M^{*2}(n,T)]^{1/2}\,. \label{mftmstar}
\ee
The single-particle potential $U(n)$ is the result of vector meson exchanges whereas the Dirac effective mass $M^*$ 
arises from scalar meson interactions.
The implementation of the above equations in the Sommerfeld expansion is made possible by the identification
\be
y = \frac{E^*}{T} \,, \quad 
\frac{dy}{dp} = \frac{p}{E^*T} \quad {\rm and} \quad
\phi(y) = \frac{E^*T}{p}g(p) .
\ee
The calculation of $d\phi/dy$, $d^3\phi/dy^3$ and $n$ proceeds as in the non-relativistic case with the replacement
$\mm \rightarrow E^*$ [cf. Eq. (\ref{genm})]. In particular, for $p_{\mu}$ we have
\be
p_{\mu} = p_F \left[1-\frac{\pi^2}{6}\frac{E_F^{*2}T^2}{p_F^4}\left(D-2+\frac{p_FE_F^{*\prime}}{E_F^*}\right)\right] \label{pmurel1}
\ee
where
\be
E_F^* = (p_F^2+M^{*2})^{1/2}      \label{ef}  \quad {\rm and} \quad
E_F^{*\prime} = \left.\frac{dE^*}{dp}\right|_{p_F} = \frac{p_F}{E_F^*}. \label{efp}
\ee
The simple dependence of $E^*$ on the momentum $p$ in Eq. (\ref{mftmstar}) leads to the correspondingly straightforward
expression (\ref{efp}) for $E_F^{*\prime}$ which, as we will show soon hereafter, results in an elementary form 
for $L_F$ and by extension the whole set of the MFT thermodynamics can be written in an uncomplicated manner. 
 
Substituting Eq. (\ref{efp}) into Eq. (\ref{pmurel1}) yields
\be
p_{\mu} = p_F \left[1-\frac{\pi^2}{6}\frac{E_F^{*2}T^2}{p_F^4}\left(D-2+\frac{p_F^2}{E_F^{*2}}\right)\right] \label{pmurel2}.
\ee
The twice-by-parts integration of Eq. (\ref{sformal}) for the entropy density in the relativistic context gives
\be
s = \frac{C_D}{T}\int dp~p^{D-1}f_p\left(E^*+\frac{p}{D}\frac{dE^*}{dp}-E^*_{\mu}\right)
\ee
where one observes the analogy with the integral term of Eq. (\ref{sd1a}). Using
\be
g_s(p) = p^{D-1}\left(E^*+\frac{p}{D}\frac{dE^*}{dp}-E^*_{\mu}\right)
\ee
we proceed as before to get the entropy density in terms of $p_{\mu}$ as
\be
s = \frac{\pi^2}{3}C_Dp_{\mu}^{D-2}E_{\mu}^*T   
 + \frac{7\pi^2}{90}C_D(D-2)(D-4)p_{\mu}^{D-6}E_{\mu}^{*3}T^3 
  \left[1+\frac{3}{(D-4)}\frac{p_{\mu}^2}{E_{\mu}^{*2}}\right]
\ee
which, with the aid of Eq. (\ref{pmurel1}), becomes
\ba
s &=& \frac{\pi^2}{3}C_Dp_F^{D-2}E_F^*T  
 + \frac{\pi^4}{45}C_D(D-2)(D-9)p_F^{D-6}E_F^{*3}T^3   \nonumber  \\
&\times& \left[1+\frac{11}{2(D-9)}\frac{p_F^2}{E_F^{*2}}-\frac{5}{2(D-2)(D-9)}\frac{p_F^4}{E_F^{*4}}\right] \,.  \label{sdenrel}
\ea
In the derivation of the last equation the weak temperature of $M^*$ in the degenerate limit
has been ignored (but not of $E_{\mu}^*$).
Combining Eq. (\ref{sdenrel}) with Eqs. (\ref{mud})-(\ref{cpd}) in $D=3$, and using the definitions [here, the Fermi temperature $T_F=p_F^2/(2E_F^*)]$
\be
a = \frac{\pi^2}{2}\frac{E_F^*}{p_F^2} \,,  \quad 
q = \frac{M^{*2}}{E_F^{*2}}\left(1-\frac{3n}{M^*}\frac{dM^*}{dn}\right)   \quad {\rm and} \quad
L_F = \frac{11}{12}\frac{p_F^2}{E_F^{*2}}-\frac{5}{12}\frac{p_F^4}{E_F^{*4}} \,,
\ee
we obtain
\ba
S &=& 2aT -\frac{16}{5\pi^2}a^3T^3(1-L_F)\,, \quad E_{th} = aT^2 - \frac{12}{5\pi^2}a^3T^4(1-L_F)  
\label{sarel3}\\
P_{th} &=& \frac{1}{3}anT^2(1+q)    
      - \frac{4}{5\pi^2}a^3nT^4\left[1-L_F+q\left(1-\frac{L_F}{3}-\frac{10}{9}\frac{p_F^4}{E_F^{*4}}\right)\right]  \label{pthrel3}\\
\mu_{th} &=& -\frac{2}{3}aT^2\left(1-\frac{q}{2}\right) 
        - \frac{4}{5\pi^2}a^3T^4q\left(1-\frac{L_F}{3}-\frac{10}{9}\frac{p_F^4}{E_F^{*4}}\right) \\
C_V &=& 2aT -\frac{48}{5\pi^2}a^3T^3(1-L_F)   \quad {\rm and} \quad
C_P = C_V + \frac{4}{9}\frac{a^2T^3(1+q)^2}{\frac{dP_0}{dn}}  \,. \label{cprel3}
\ea
As in the nonrelativistic case, when conditions are such that $dP_0/dn$ is small, one must use derivatives of the pressure 
with respect to $n$ and $T$ that include thermal contributions to $\mathcal{O}(T^4)$ in the calculation of $C_P$. Explicitly,
\ba
\left(\left.\frac{\partial P_{th}}{\partial T}\right|_n\right)^2 &=& \frac{4}{9}a^2n^2T^2(1+q)^2 \nonumber \\
  &\times& \left\{1-\frac{48}{5\pi^2}\frac{a^2T^2}{1+q}\left[1-L_F+q\left(1-\frac{L_F}{3}-\frac{10}{9}\frac{p_F^4}{E_F^{*4}}\right)\right]\right\} \\
\left.\frac{\partial P}{\partial n}\right|_T  &=& \frac{dP_0}{dn} + \frac{T^2a}{9}\left[(1+q)(2-q)+3n\frac{dq}{dn}\right]  \nonumber \\
 &+& \frac{4}{5\pi^2}T^4a^3q\left[1-L_F+\left(q-\frac{n}{q}\frac{dq}{dn}\right)\left(1-\frac{L_F}{3}-\frac{10}{9}\frac{p_F^4}{E_F^{*4}}\right)
               \right. \nonumber   \\
 &+& \left.\frac{n}{q}\frac{dL_F}{dn}\left(1+\frac{q}{3}\right)+\frac{40}{27}\frac{p_F^4}{E_F^{*4}}q\right]  \\
\frac{dq}{dn} &=& 
 -\frac{2}{3n}q(1-q)+\frac{1}{M^*}\frac{dM^*}{dn}q-\frac{M^*}{2E_F^{*2}}\left(\frac{dM^*}{dn}+\frac{3n}{2}\frac{d^2M^*}{dn^2}\right).
\ea

\section{RESULTS AND DISCUSSION}
\label{Sec:Results}

Here, we compare the results from FLT and FLT+NLO  with the exact numerical results 
for the two non-relativistic models (MDI(A) and SkO$^\prime$), and for a  relativistic 
mean-field theoretical model (MFT). 
Numerical techniques for obtaining the exact  numerical results are detailed in Refs. \cite{jel,APRppr}.
The thermal properties presented for PNM and SNM are at a temperature of 
$T=20$ MeV. \\

\begin{figure*}[!htb]
\centering
\begin{minipage}[b]{\linewidth}
\centering
\includegraphics[width=12cm]{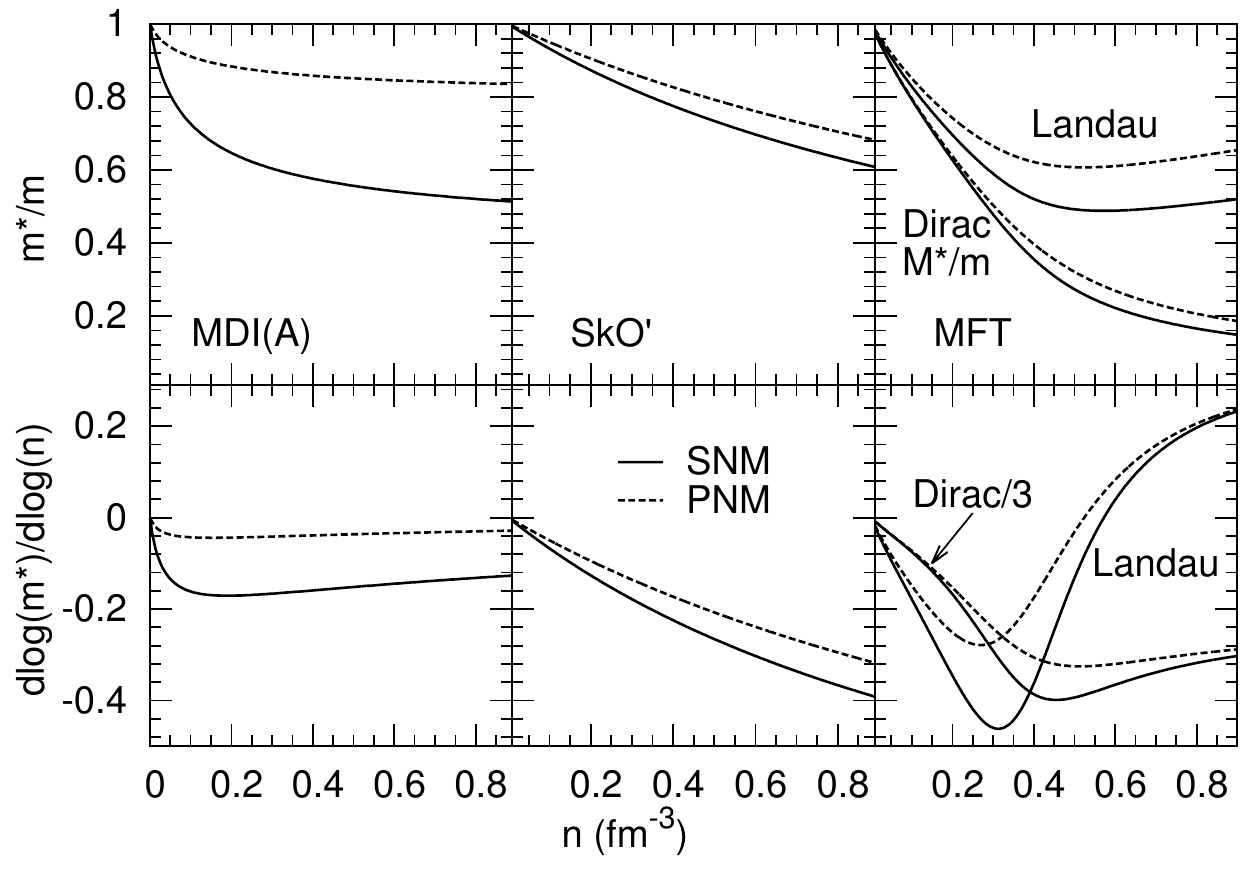}
\end{minipage}
\vskip -0.5cm
\caption{Top panels: Effective mass vs density for non-relativistic potential models (MDI(a) and SkO$^\prime$) and relativistic mean-field theoretical model (MFT) for symmetric nuclear matter (SNM) and pure neutron matter (PNM). Bottom panels:  
Logarithmic derivative of the effective mass with respect to density vs density.}
\label{3mod_Ms_dMs}
\end{figure*}

In the top panels of Fig. \ref{3mod_Ms_dMs},  the Landau effective masses $m_n^*$ of the neutron scaled with its vacuum value 
are shown as a function of baryon density $n$. For the MFT model, both $m_n^*= E_{F_{n}}^*= {\sqrt{M^{*^{2}} + k_{F_{n}}^{2}}}$ and the  
Dirac effective mass $M^*$ are shown. Noteworthy points for the non-relativistic models are: (i)  The isospin splittings are  qualitatively 
similar - $m_n^*/m$ being larger for PNM than for SNM - although quantitative differences are present, and (ii) except for $n$ up 
to $0.2~{\rm fm}^{-3}$, the decrease with increasing $n$  for the MDI model is relatively slow (logarithmic decline) compared with that 
for the SkO$^\prime$ model [$(1+\beta n)^{-1}$ fall off]. This overall flatness of $m^*$ for the MDI model is a direct consequence of the 
momentum structure of its single-particle potential which causes it to saturate at high momenta. For the MFT model,  $M^*$ decreases 
monotonically with $n$ to values lower than those for the non-relativistic models. The Landau mass $m^*$, however, exhibits a non-monotonic 
behavior, attaining a minimum for $n_{min}=0.57~(0.52)~{\rm fm}^{-3}$ obtained from the solution of 
\be
\frac {p_F}{M^*} + \frac {dM^*}{dp_F} = 0
\ee
for the case of  SNM (PNM), and increasing thereafter due to the 
monotonic increase of $k_{F_{n}}$ with $n$. Physically, $n_{min}$ marks the transition of nucleons well into the relativistic region. The density $n_R$ at which $p_F=M^*$ occurs at 
\be
n_R = 0.643~\gamma \left(\frac {M^*}{m}\right)^3{\rm fm}^{-3}\,
\ee
about $(2/3)~n_{min}=0.38~(0.34)~{\rm fm}^{-3}$ for SNM (PNM) which signals the onset of relativistic effects which become progressively important for $n \geq n_R$.\\

The bottom panels of Fig. \ref{3mod_Ms_dMs} show the logarithmic derivatives of $m_n^*$
vs $n$. Also shown for the MFT model is $d \ln M^*/d \ln n$ which has been divided by a factor of 3
to fit within the figure. The logarithmic derivatives $m^*$ for MDI(A) show little variation with $n$ at supra-nuclear densities. 
In contrast, results for the SkO$^\prime$ model, which take the simple form $(m^*_n/m)-1$, 
show a significant variation with $n$.  The logarithmic derivative of $M^*$ in 
MFT drops to values considerably lower than for the other two models. This derivative remains negative but
approaches a constant value at large densities. The logarithmic derivative of the 
Landau effective mass drops until $\sim~2n_0$, but then increases to positive values. This behavior is 
a reflection of the minimum that occurs for $m_n^*$ in this model. As will be seen below, the density dependences of the 
effective masses and their logarithmic derivatives determine the behavior of all the thermal properties in FLT. Higher 
order derivatives of the Landau effective mass function in Eq. (\ref{genm}) appear in FLT+NLO. \\

\begin{figure*}[!htb]
\centering
\begin{minipage}[b]{\linewidth}
\centering
\includegraphics[width=12cm]{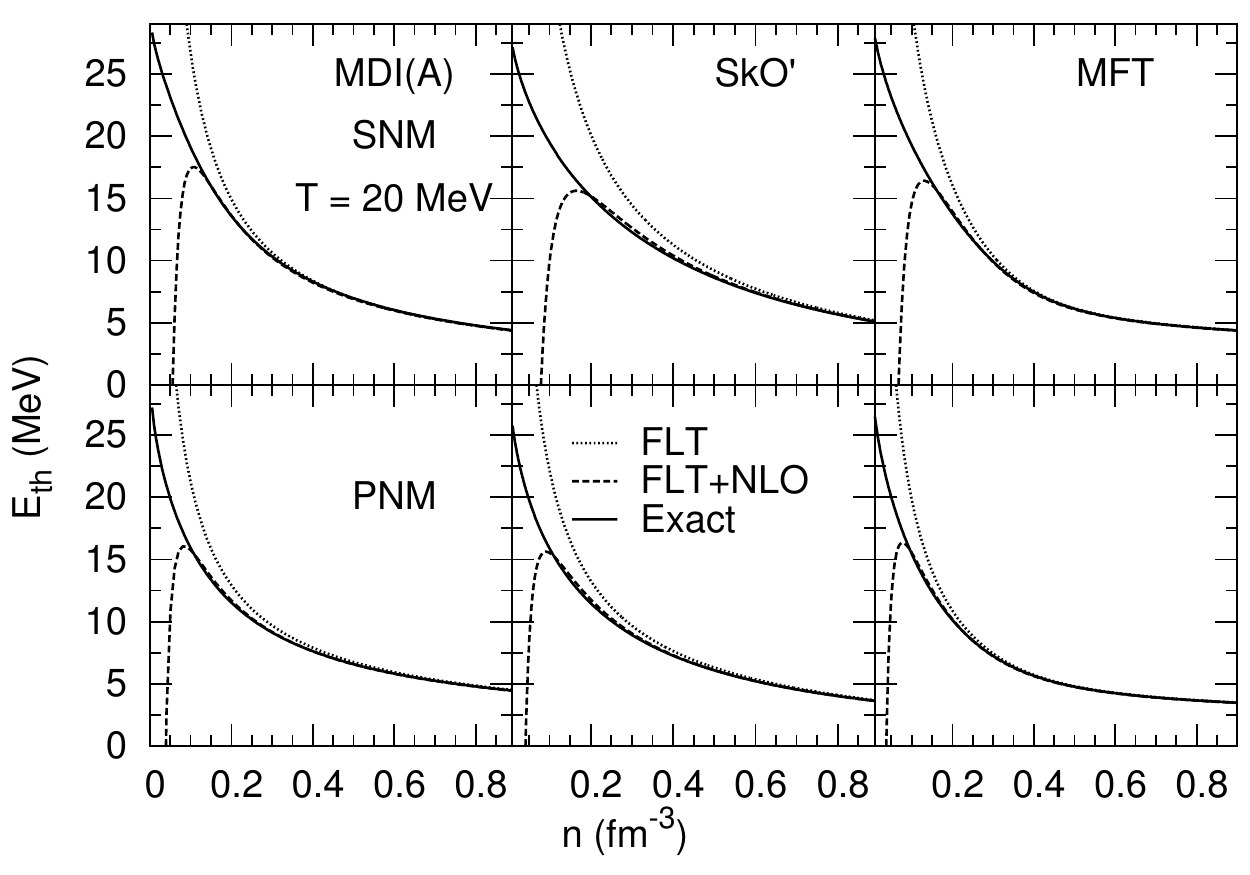}
\end{minipage}
\vskip -0.5cm
\caption{Thermal energy vs baryon density for the three models
at a temperature of  $T=20$ MeV. Results for SNM are 
in the top panels and for PNM in the bottom panels.}
\label{3mod_Eth}
\end{figure*}

The FLT and FLT+NLO results for the thermal energy, $E_{th}$ vs $n$,  are compared with the exact numerical results in 
Fig. \ref{3mod_Eth}. The NLO corrections to FLT yield agreement with the exact results down to sub-nuclear densities of 
0.5 to 1 $n_0$ compared to 2-3~$ n_0$ for FLT. As is the case with FLT, slightly but systematically better agreement with 
FLT+NLO occurs for PNM than for SNM for all the thermodynamic quantities we study. This 
is a consequence of the fact that the neutron density in PNM ($n_n=n$) is twice the neutron density 
in SNM ($n_n=n/2$); PNM is more degenerate than SNM at the same baryon density $n$.      \\

\begin{figure*}[!htb]
\centering
\begin{minipage}[b]{\linewidth}
\centering
\includegraphics[width=12cm]{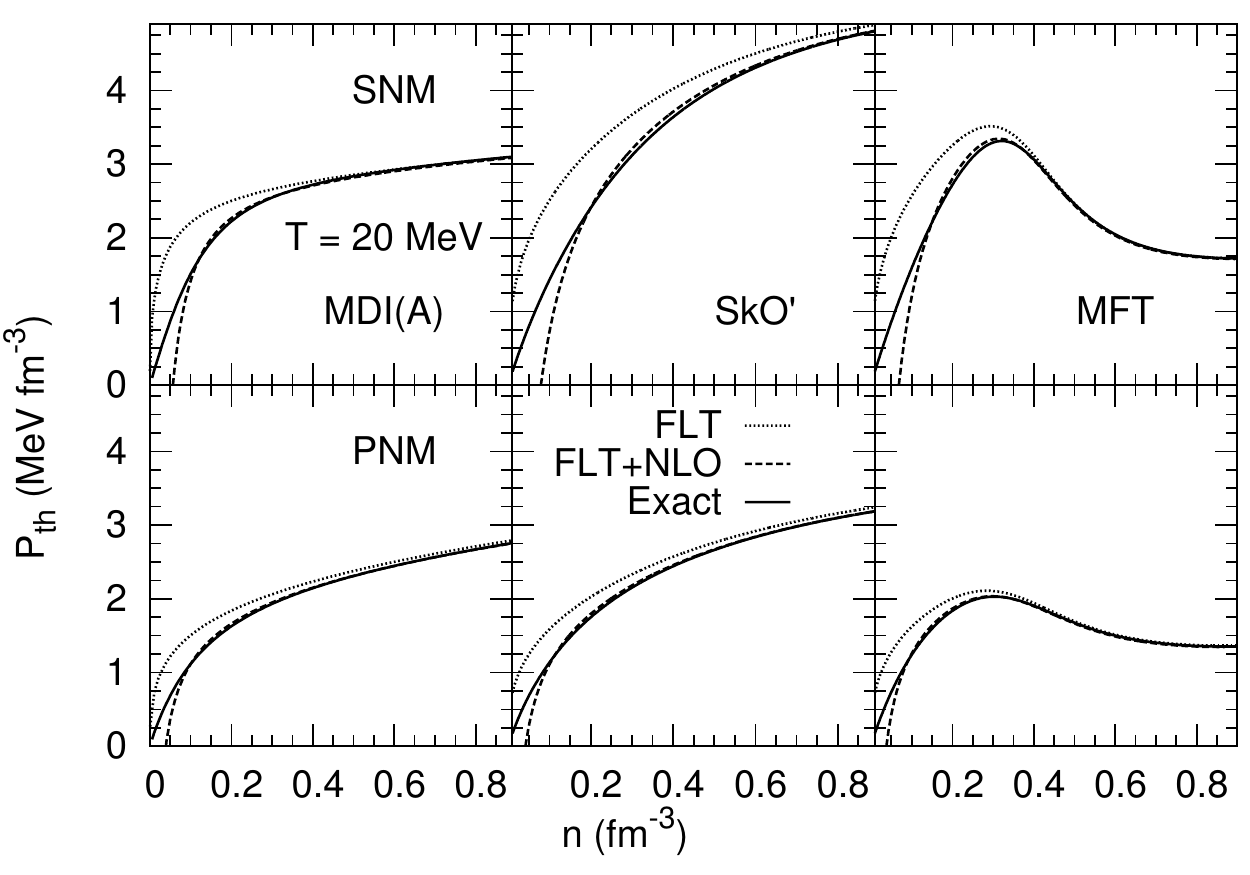}
\end{minipage}
\vskip -0.5cm
\caption{Same as Fig. \ref{3mod_Eth}, but for thermal pressure.}
\label{3mod_Pth}
\end{figure*}

In Figure \ref{3mod_Pth}, we show the convergence of FLT and FLT+NLO results to the exact numerical results for $P_{th}$. 
For all three models, the FLT+NLO results extend the agreement with the exact results to lower densities than those of FLT.  
The SkO$^\prime$ model shows the greatest changes relative to FLT for both SNM and PNM. 
Both the MDI(A) and MFT results also show improvement using FLT+NLO with better agreement occurring 
for PNM. For $P_{th}$ at $n > n_0$, the influence of  $m^*(n)$ and its logarithmic derivative with respect to $n$ (in FLT), and the higher derivatives of ${\cal M}$ (in FLT+NLO) are amply demonstrated: (i) for MDI(A), $P_{th}$  grows very slowly with $n$; 
(ii) considerably larger growth with $n$ is exhibited for SkO$^\prime$ than for MDI(A); and (iii) for MFT, the prominent peaks in $P_{th}$ at intermediate densities are due to the minima in $m^*$ and $d \ln M*/d \ln n$ at similar densities. 
For asymptotic densities, $P_{th} \propto n^{4/3}$ in MFT models characteristic of massless particles. 
A corresponding behavior is also present in the thermal chemical potentials (see below).\\
 
\begin{figure*}[!htb]
\centering
\begin{minipage}[b]{\linewidth}
\centering
\includegraphics[width=12cm]{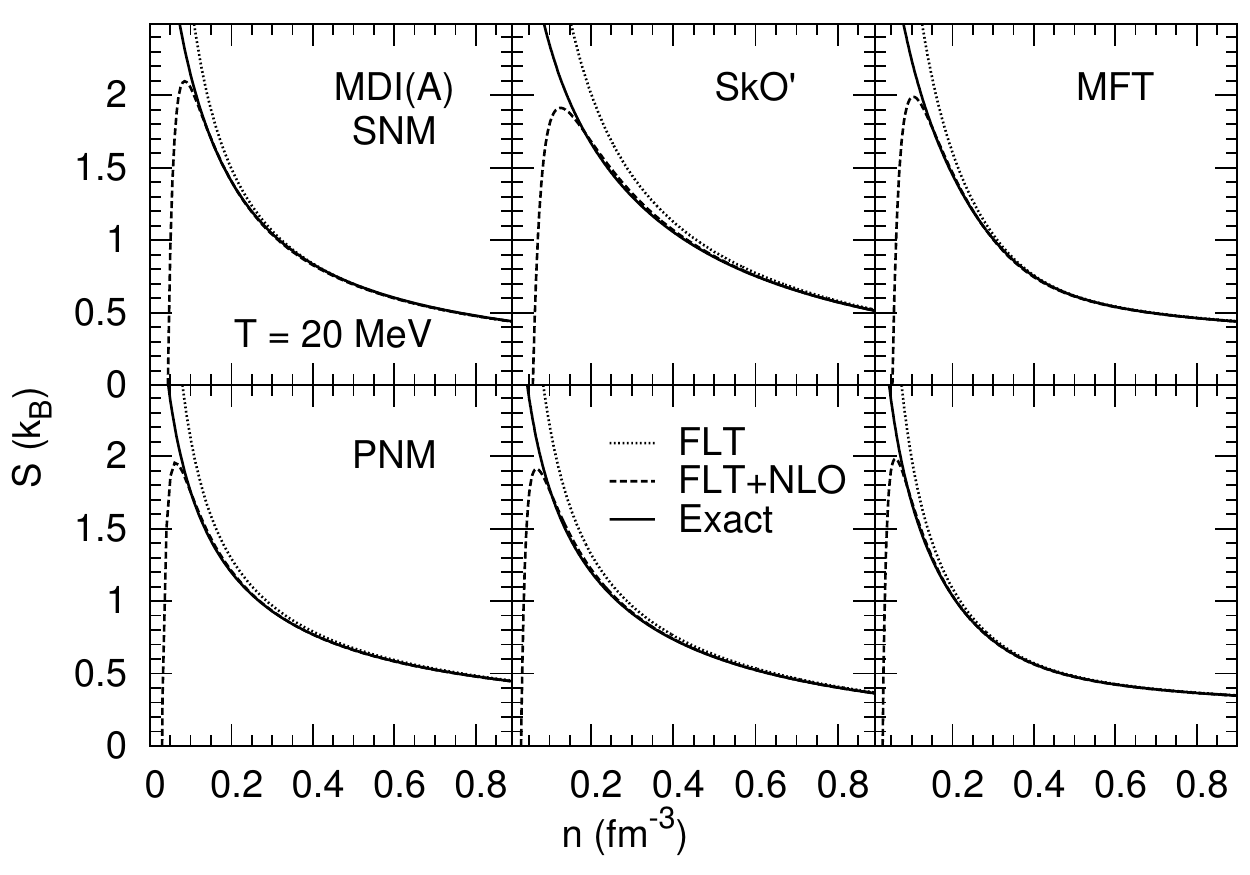}
\end{minipage}
\vskip -0.5cm
\caption{Same as Fig. \ref{3mod_Eth}, but for entropy per baryon.}
\label{3mod_S}
\end{figure*}

Comparisons of the entropy per baryon, $S$, are shown in Fig. \ref{3mod_S}.  
For all three models, the agreement between the exact results and those of FLT+NLO 
extends to below $n_0$, and to as low as $n_0/4$ for PNM with MFT. This is an improvement from the FLT 
results for which convergence ranged from $1.5n_0$ (FLT and PNM) to $3n_0$ 
(SkO$^\prime$ and SNM). \\ 

\begin{figure*}[!htb]
\centering
\begin{minipage}[b]{\linewidth}
\centering
\includegraphics[width=12cm]{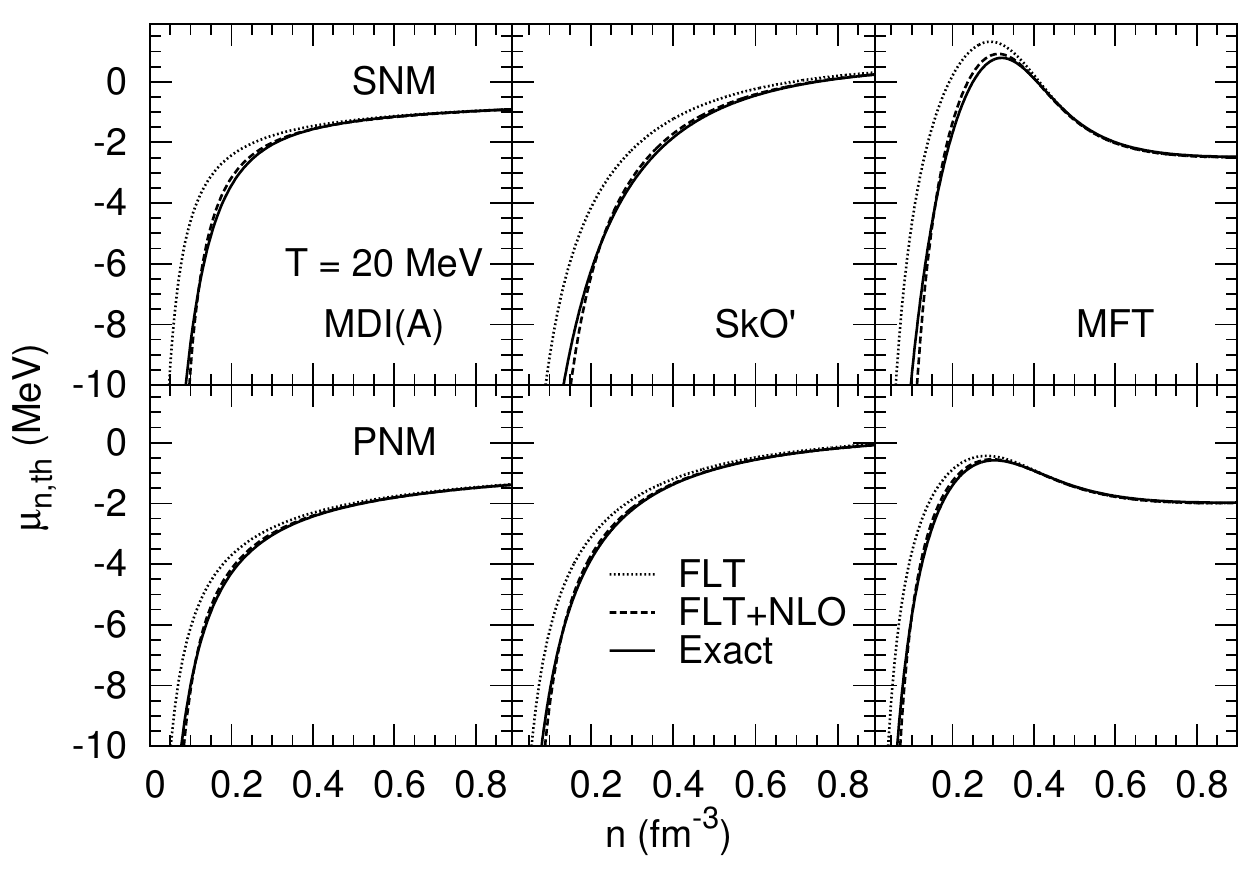}
\end{minipage}
\vskip -0.5cm
\caption{Same as Fig. \ref{3mod_Eth}, but for thermal neutron chemical potential.}
\label{3mod_MuNth}
\end{figure*}

In Fig. \ref{3mod_MuNth},  the thermal parts of the neutron chemical potentials, $\mu_{n,th}$, are shown as a function of $n$.
Results from FLT+NLO lie closer to the exact results than do those of FLT with the agreement extending to sub-nuclear densities.  
As with the other state variables, the agreement is quantitatively better for PNM than for SNM.  \\

\begin{figure*}[!htb]
\centering
\begin{minipage}[b]{\linewidth}
\centering
\includegraphics[width=12cm]{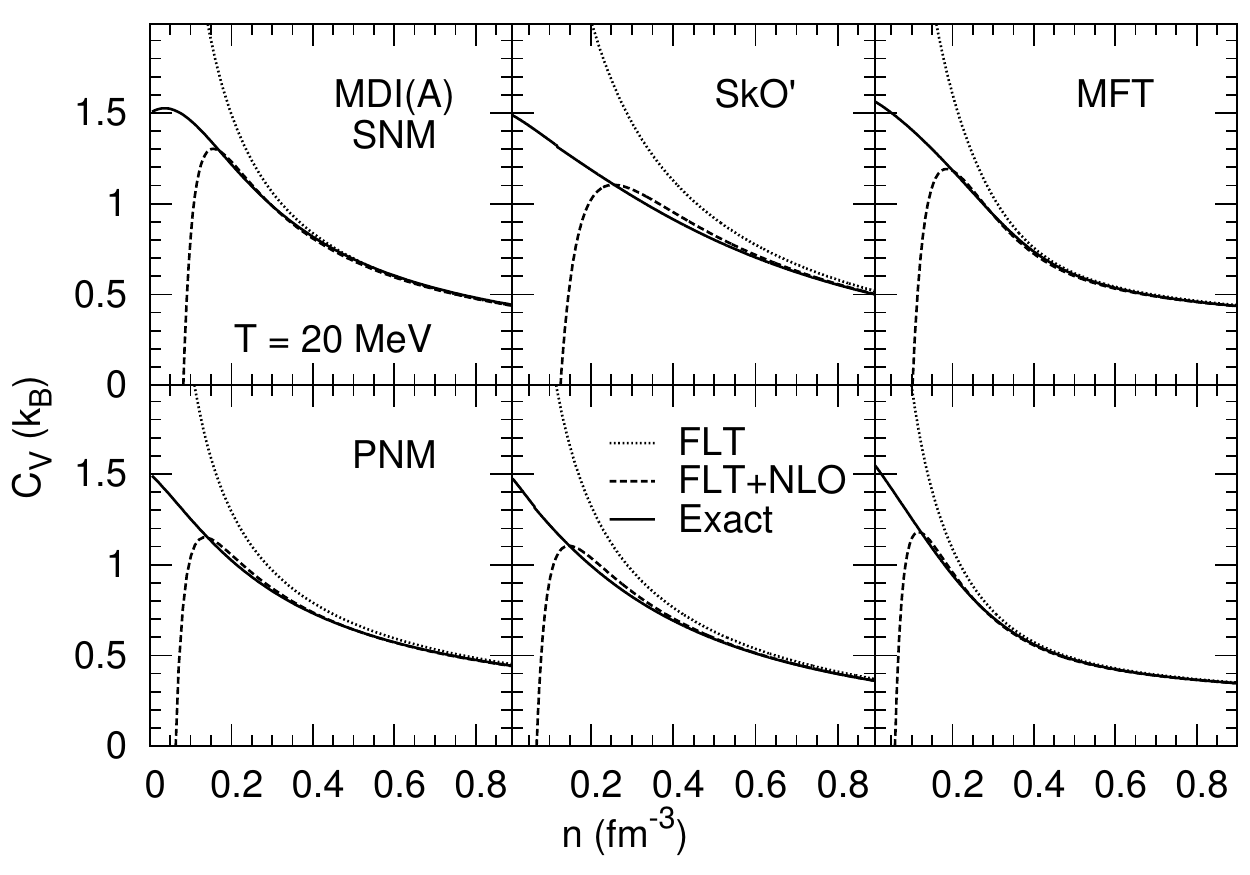}
\end{minipage}
\vskip -0.5cm
\caption{Same as Fig. \ref{3mod_Eth}, but for specific heat at constant volume.} 
\label{3mod_Cv}
\end{figure*}

The specific heat at constant volume, $C_V$,  is presented in Fig. \ref{3mod_Cv} as a function of $n$.
For all models, the FLT+NLO results show better agreement with the exact results than those of FLT.
The lowest density for which the agreement extends differs between the models, that for the 
SkO$^\prime$ model being higher than for the other two models particularly for SNM. \\

\begin{figure*}[!htb]
\centering
\begin{minipage}[b]{\linewidth}
\centering
\includegraphics[width=12cm]{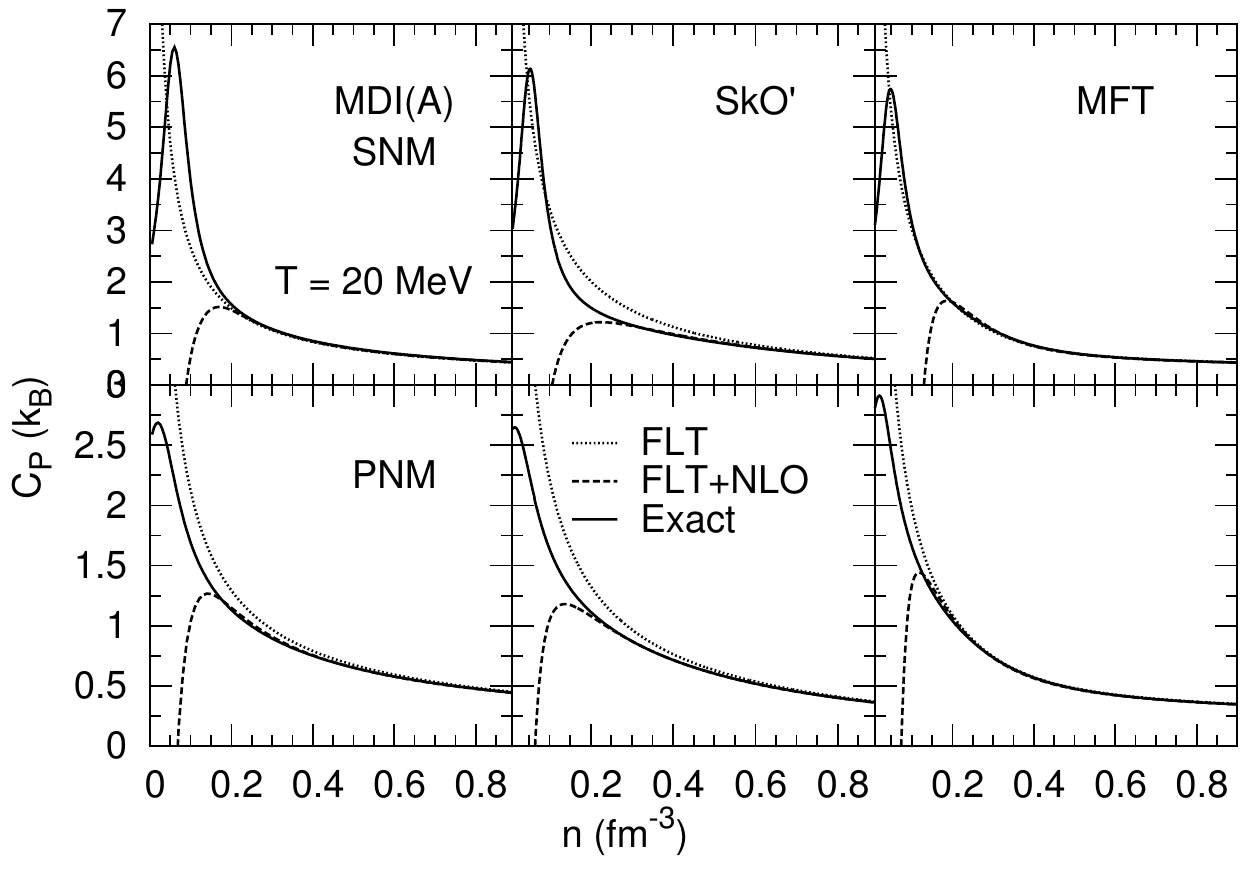}
\end{minipage}
\vskip -0.5cm
\caption{Same as Fig. \ref{3mod_Eth}, but for specific heat at constant pressure.} 
\label{3mod_Cp}
\end{figure*}

In Figure \ref{3mod_Cp} we show the specific heat at constant pressure and its limiting cases as 
functions of $n$. The maxima at low sub-nuclear densities in the exact numerical results are related to 
the liquid-gas phase transition of nucleonic matter which occurs at $T \sim$ 15-20 MeV. 
The convergence between the exact and approximate results for MDI(A) and MFT follows a pattern similar to the quantities discussed previously with the conspicuous exception of FLT  which appears to outperform FLT+NLO for SNM. We attribute this feature to a numerical accident, possibly due to the proximity to the phase transition.  Note that FLT 
begins to deviate from the exact result for densities below $n \sim 3n_0$ in the case of SkO$^\prime$. 
As our analysis here is concerned with the degenerate region, this failure at sub-nuclear densities is not surprising. For an adequate treatment  in the non-degenerate region, see Ref. \cite{cons15}. \\

\begin{figure*}[!htb]
\centering
\begin{minipage}[b]{\linewidth}
\centering
\includegraphics[width=12cm]{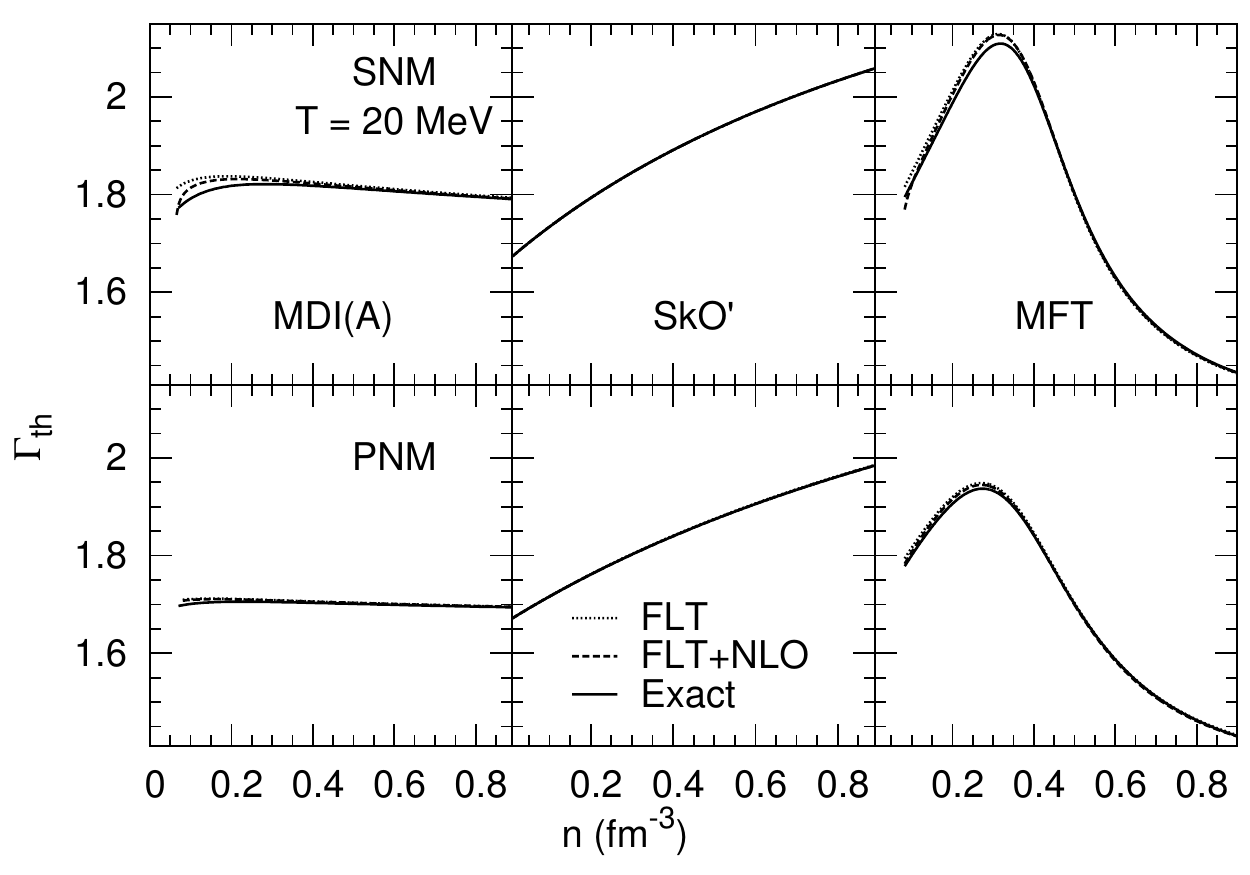}
\end{minipage}
\vskip -0.5cm
\caption{Same as Fig. \ref{3mod_Eth}, but for the thermal index.} 
\label{3mod_Gth}
\end{figure*}

The thermal index,  $\Gamma_{th}=1+(P_{th}/\varepsilon_{th})$, is shown in Fig. \ref{3mod_Gth}. For the SkO$^{\prime}$ 
model, the FLT result is exact for all regions of degeneracy. This happenstance is due to the fact that for non-relativistic nucleons with only contact interactions, $P_{th}$ and $\varepsilon_{th}$ can be written entirely in terms of their ideal-gas counterparts as
\ba
P_{th}(n,T) &=& P_{th}^{id}(n,T;m^*)~Q \nonumber \\
\varepsilon_{th}(n,T) &=& \varepsilon_{th}^{id}(n,T;m^*)\,, \qquad
\frac {P_{th}^{id}}{\varepsilon_{th}^{id}} = \frac 23 \,
\ea
regardless of the degree of degeneracy.
The $n$-dependence of $m^*$ for Skyrme-like interactions thus yields
\be
\Gamma_{th}= \frac{8}{3} - \frac{m^*(n)}{m}\,.
\ee
For the finite-range MDI(A) and the relativistic MFT models, the FLT+NLO results mildly improve the FLT results in reproducing the exact ones. This marginal improvement is related to the ratio $P_{th}/\varepsilon_{th}$ in $\Gamma_{th}$ in these cases, which amount to a constant plus a correction due to effects of $n$ plus a second correction due to $(n,T)$ effects,
which means that temperature effects are sub-leading, and therefore very weak. Note that the results for the three models differ significantly from each other, both qualitatively and quantitatively. These differences are due to the differences in the effective mass functions of the models.  

In physical applications involving neutron stars and supernovae, contributions from leptons (electrons and muons) and photons 
must be included to all of the state variables.  Thermal effects from these sources are adequately given by their free gas forms, and numerical methods for their calculation for arbitrary degeneracy can be found in Ref. \cite{jel}. The influence of thermal effects from leptons and photons to the total has been detailed in Ref. \cite{cons15}, and will not be repeated here.

\section{SUMMARY AND CONCLUSIONS}
\label{Sec:Conclusions}

For homogeneous systems of fermions in the limit of extreme degeneracy ($T/T_F \ll 1$, where $T_F$ is the Fermi temperature), Landau's Fermi Liquid Theory (FLT) provides simple analytical expressions that are model independent for the thermal state variables ({\em e.g.,} entropy, energy, pressure, chemical potential, and specific heats) \cite{flt}.  In the absence of collective 
excitations close to the Fermi surface, thermal effects are primarily determined by the nucleon's Landau 
effective mass and its first density derivative which in turn depend on the momentum-dependence of the $T=0$ single-particle energy spectrum.  

In this work, we have developed a method by which thermal effects in near-degenerate to degenerate matter can be described to next-to-leading order in $T/T_F$ for models with general momentum dependences in their single-particle potentials. 
Analytical formulas valid to next-to-leading order in $T/T_F$ for all of the thermal state variables are presented. 
The entropy density and specific heats are carried to ${\cal O} (T/T_F)^3$ whereas the energy density and pressure to 
${\cal O} (T/T_F)^4$, extending the leading order results of FLT.  These extensions involved the use of a generalized Landau effective mass function which enables the calculation of the entropy density, and thereafter the other state variables, for a general single-particle spectrum. In special cases, {\em e.g.,} models with contact interactions, knowledge of the Landau effective mass suffices. 
In the case of finite-range interactions, momentum derivatives of the Landau effective mass function
up to second order are required to satisfy the thermodynamic identity. Our results are valid for potential and field-theoretical models 
as long as the underlying interactions yield a single-particle spectrum that is weakly dependent on temperature in the degenerate limit. 

We find that ${\cal O}(T^3)$  corrections in 2-dimensions for non-relativistic models appear only if they include 
finite-range interactions or, equivalently, if the momentum content of their mean field is something other than quadratic. These contributions supplement the ${\cal O}(T^2)$ non-analytic contributions even with zero-range interactions established previously \cite{CB93}.

To illustrate the density region of their applicability, numerical results from the new formulas were compared with those of 
exact numerical calculations for zero- and finite-range potential models as well as for relativistic mean-field  theoretical 
(MFT) models widely used in astrophysical applications of hot and dense nuclear matter. In all cases, excellent agreement with the exact results was found even to sub-nuclear densities of 
$\sim 0.1~{\rm fm}^{-3}$ for $T=20$ MeV, whereas FLT results are valid only for densities beyond   $\sim 0.3~{\rm fm}^{-3}$.  
(For low temperatures and below $\sim 0.1~{\rm fm}^{-3}$, inhomogeneous phases with nuclei and pasta-like configurations are known to exist, and must be treated separately.)  
In addition to providing physical insights, our analytical 
results facilitate a rapid evaluation of the EOS in the homogeneous phase (important for computer-time consuming large-scale 
simulations of supernovae, proto-neutron stars, and mergers of binary compact objects). Our formulas can be used for 
any $T=0$ quasi-particle spectrum, {\em e.g.,}  those extracted from Brueckner-Hartree-Fock and Dirac-Brueckner-Hartree-Fock approaches, extensions of MFT models  with non-linear derivatives or with 2-loop effects, and  effective field-theoretical approaches. 

Examples of contributions  not included in our work arise from, {\em e.g.,} non-analytic contributions from single  particle-hole excitations and, collective and paring correlations close to the Fermi surface.  Their roles as functions of densities and temperatures of relevance to astrophysical phenomena need further investigation.

\section*{ACKNOWLEDGEMENTS}
This work was supported by the U.S. DOE under Grants No. DE-FG02-93ER-40756 
and No. DE-FG02-87ER-40317.

\newpage

%
%

\clearpage   

\appendix
\section{RESULTS RELATED TO $L_F$}
\label{Sec:App.A}

Here we collect relations required to calculate the quantity $L_F$ defined in Eq. (\ref{LF}) of Sec. \ref{Sec:D3}. 

\subsection*{Model-independent relations}
\ba
\mm' &\equiv& \frac{d\mm}{dp} = \frac{\mm^2}{p^2}\left(\frac{dR}{dp}-p\frac{d^2R}{dp^2}\right) \\
\mm''&=& \frac{2\mm^3}{p^4}\left(\frac{dR}{dp}-p\frac{d^2R}{dp^2}\right)
  -\frac{2\mm^2}{p^3}\left(\frac{dR}{dp}-p\frac{d^2R}{dp^2}+\frac{p^2}{2}\frac{d^3R}{dp^3}\right)  \\
&=& -2\frac{\mm'}{p}\left(1-\frac{\mm}{p}\right)-\frac{\mm^2}{p}\frac{d^3R}{dp^3}  \equiv \frac{d^2\mm}{dp^2} \\
\frac{dL_F}{dn} &=& \frac{p_F}{3n}\frac{dL_F}{dp_F}  \\
\frac{dL_F}{dp_F} &=& \frac{5}{12}\frac{\mm_F'}{m^*}\left(1-p_F\frac{\mm_F'}{m^*}\right)
\left(1-\frac{42}{5}p_F\frac{\mm_F'}{m^*}\right)  \nonumber \\
&-& \frac{m^*}{6}\left.\frac{d^3R}{dp^3}\right|_{p_F}\left(1+\frac{49}{2}p_F\frac{\mm_F'}{m^*}\right)
 - \frac{7}{12}p_Fm^*\frac{d}{dp_F}\left.\frac{d^3R}{dp^3}\right|_{p_F}  
\ea
The subscript $F$ denotes evaluation at $p=p_F$.

\subsection*{MDI model-specific realations}
\ba
\left.\frac{dR}{dp}\right|_{p_F} &=& \frac{C_{\gamma}}{n_0}\frac{\Lambda^2}{\pi^2\hbar^3}
\left[1-\frac{1}{2}\left(1+\frac{\Lambda^2}{2p_F^2}\right)\ln\left(1+\frac{4p_F^2}{\Lambda^2}\right)\right]    \\
\left.\frac{d^2R}{dp^2}\right|_{p_F} &=& -\frac{2C_{\gamma}}{n_0}\frac{\Lambda^2}{\pi^2\hbar^3}\frac{1}{p_F(4p_F^2+\Lambda^2)}
\left[(3p_F^2+\Lambda^2)\frac{}{}\right.  \nonumber \\
  &-&\left.(p_F^2+\Lambda^2)\left(1+\frac{\Lambda^2}{4p_F^2}\right)\ln\left(1+\frac{4p_F^2}{\Lambda^2}\right)\right]  \\
\left.\frac{d^3R}{dp^3}\right|_{p_F} &=& \frac{2C_{\gamma}}{n_0}\frac{1}{\pi^2\hbar^3}\frac{1}{p_F^2(4p_F^2+\Lambda^2)^2} \nonumber \\
&\times& \left[8p_F^6+34p_F^4\Lambda^2+21p_F^2\Lambda^4+3\Lambda^6\frac{}{}\right.  \nonumber \\
  &-& \left.\frac{3\Lambda^2}{4}\left(1+\frac{\Lambda^2}{p_F^2}\right)(4p_F^2+\Lambda^2)^2\ln\left(1+\frac{4p_F^2}{\Lambda^2}\right)\right] \\
\frac{d}{dp_F}\left.\frac{d^3R}{dp^3}\right|_{p_F} &=& -\frac{2C_{\gamma}}{n_0}\frac{1}{\pi^2\hbar^3}\frac{1}{p_F^5(4p_F^2+\Lambda^2)^3}
\nonumber \\
&\times& \left[4p_F^2(24p_F^8+134p_F^6\Lambda^2\frac{}{}
 + 118p_F^4\Lambda^4+33p_F^2\Lambda^6+3\Lambda^8)\right. \nonumber \\
  &-& \left.3\Lambda^2(p_F^2+\Lambda^2)(4p_F^2+\Lambda^2)^3\ln\left(1+\frac{4p_F^2}{\Lambda^2}\right)\right]
\ea

\section{MDI HAMILTONIAN DENSITY}
\label{Sec:App.B}

For the MDI models, the Hamiltonian density is composed of 
terms arising from  kinetic sources, $\mathcal{H}_k$,  density-dependent  
interactions, $\mathcal{H}_d$, and  momentum-dependent interactions, $\mathcal{H}_m$:  
\ba
\mathcal{H} =\mathcal{H}_k+\mathcal{H}_d+\mathcal{H}_m \,.
\ea
At $T=0$, 
\ba
\mathcal{H}_k &=& \frac{1}{2m}(\tau_n+\tau_p)=\frac{1}{2m}\frac{1}{5\pi^2\hbar^3}(p_{Fn}^5+p_{Fp}^5)  \label{MDYI_H_0T_0}  \\
\mathcal{H}_d &=& \frac{A_1}{2n_0}n^2+\frac{A_2}{2n_0}n^2(1-2x)^2   
    +\frac{B}{\sigma+1}\frac{n^{\sigma+1}}{n_0^{\sigma}}\left[1-y(1-2x)^2\right]  \label{MDYI_H_0T_d}\\
\mathcal{H}_m &=& \frac{C_l}{n_0}(I_{nn}+I_{pp})+\frac{2C_u}{n_0}I_{np}   
\label{MDYI_H_0T}
\ea
with
\ba
x &=& n_p/n \,,  \qquad 
p_{Fi} = (3\pi^2n_i\hbar^3)^{1/3} \\
I_{ij} &=& \frac{8\pi^2\Lambda^2}{(2\pi\hbar)^6}\left\{p_{Fi}p_{Fj}(p_{Fi}^2+p_{Fj}^2)-\frac{p_{Fi}p_{Fj}\Lambda^2}{3}\right.  \nonumber \\
      &+& \frac{4\Lambda}{3}(p_{Fi}^3-p_{Fj}^3)\arctan\left(\frac{p_{Fi}-p_{Fj}}{\Lambda}\right)  \nonumber  \\
      &-& \frac{4\Lambda}{3}(p_{Fi}^3+p_{Fj}^3)\arctan\left(\frac{p_{Fi}+p_{Fj}}{\Lambda}\right)   \nonumber \\
      &+& \left[\frac{\Lambda^4}{12}+\frac{(p_{Fi}^2+p_{Fj}^2)\Lambda^2}{2}-\frac{(p_{Fi}^2-p_{Fj}^2)^2}{4}\right] \nonumber \\
      &\times& \left.\ln\left[\frac{(p_{Fi}+p_{Fj})^2+\Lambda^2}{(p_{Fi}-p_{Fj})^2+\Lambda^2}\right]\right\}\,.
      \label{Iij}
\ea
In this work we use the coefficients $A_1=-69.48$ MeV, $A_2=-29.22$ MeV, $B=100.1$ MeV, $\sigma=1.362$, $y=-0.0328$,
$C_l=-23.06$ MeV, $C_u=-105.9$ MeV, and $\Lambda=420.9$ MeV \cite{cons15}.

When a calculation for multiple-species is undertaken, highly asymmetric configurations should
be avoided as the various particle types involved will be in different regimes of degeneracy thus resulting in a slower convergence
relative to the single-species case. Furthermore, one should also refrain from using multiple species results with $x=0$ or 1 as
numerical complications arise; namely, division by 0 occurs in the level density parameters $a_i$.

\section{MEAN-FIELD THEORETICAL MODEL  (MFT)}
\label{Sec:App.C}

The MFT model used here involves the exchange of $\sigma$, $\omega$ and $\rho$ mesons (scalar, vector and
 iso-vector, respectively) \cite{Muller96}. Its Lagrangian density is 
\ba
\mathcal{L} &=& \bar \Psi\left[i\gamma_{\mu}\partial^{\mu}-\gamma_0g_{\omega}\omega_0-\gamma_0\frac{g_{\rho}}{2}\rho_0\tau_3
  -(M-g_{\sigma}\sigma_0)\right]\Psi  \nonumber  \\
&-& \frac{1}{2}\left[m_{\sigma}^2\sigma_0^2+\frac{\kappa}{3}(g_{\sigma}\sigma_0)^3+\frac{\lambda}{12}(g_{\sigma}\sigma_0)^4\right]
 + \frac{1}{2}m_{\omega}^2\omega_0^2+\frac{1}{2}m_{\rho}^2\rho_0^2
\ea
which yields the following meson equations of motion:
\ba
g_{\sigma}\langle\bar \Psi \Psi\rangle &=& g_{\sigma}n_S  
 = m_{\sigma}^2\sigma_0 + \frac{\kappa}{2}g_{\sigma}^3\sigma_0^2 + \frac{\lambda}{6}g_{\sigma}^4\sigma_0^3  \\
\omega_0 &=& \frac{g_{\omega}}{m_{\omega}^2}\langle \Psi^{\dag}\Psi\rangle = \frac{g_{\omega}}{m_{\omega}^2}n  \\
\rho_0 &=& \frac{g_{\rho}}{2m_{\rho}^2}\langle \Psi^{\dag}\tau_3\Psi \rangle = \frac{g_{\rho}}{2m_{\rho}^2}(n_n-n_p)
\ea
in the mean-field approximation with classical expectation values denoted by the subscript $``0"$.
The equation of motion for the nucleon ($\Psi$) field is
\be
\left[i\gamma_{\mu}\partial^{\mu}-\gamma_0(g_{\omega}\omega_0+\frac{g_{\rho}}{2}\rho_0\tau_3)-M^*\right]\Psi = 0 \label{neom}
\ee
where $M^*=M-g_{\sigma}\sigma_0$, and yields the nucleon single-particle energy spectrum
\be
\epsilon_{i\pm} = \pm E_i^*+\frac{g_{\omega}^2}{m_{\omega}^2}n + \frac{g_{\rho}^2}{4m_{\rho}^2}(n_i-n_j) \,, \quad 
E_i^* = (p_i^2 + M^{*2})^{1/2} \,.
\ee
The subscripts $i$, $j$ refer to the nucleon species, the positive sign to the particles and the negative sign to the antiparticles.
The thermodynamics of the system are obtained from its energy-momentum tensor
\ba
T_{\mu\nu} &=& \frac{\partial\mathcal{L}}{\partial(\partial_{\mu}\phi)}\partial_{\nu}\phi-g_{\mu\nu}\mathcal{L} \\
  &=& i\bar\Psi\gamma_{\mu}\partial_{\nu}\Psi+\frac{g_{\mu\nu}}{2}
             \left[m_{\sigma}^2 \sigma_0^2 + \frac{\kappa}{3}(g_{\sigma}\sigma_0)^3
  +\frac{\lambda}{12}(g_{\sigma}\sigma_0)^4-m_{\omega}^2 \omega_0^2-m_{\rho}^2\rho_0^2\right]. \nonumber \\
\ea
For an isotropic system in its rest-frame, the energy density and the pressure are given by the diagonal elements
of $T_{\mu\nu}$ as
\ba
\varepsilon &=& \langle T_{00} \rangle  
 = 2\sum_i\int f_{p_i}(p_i^2 + M^{*2})^{1/2}\frac{d^3p_i}{(2\pi\hbar)^3}+\frac{g_{\omega}^2}{2m_{\omega}^2}n^2   \nonumber \\
&+&\frac{g_{\rho}^2}{8m_{\rho}^2}(n_p-n_n)^2 +\frac{1}{2}\left[m_{\sigma}^2 \sigma_0^2 + \frac{\kappa}{3}(g_{\sigma}\sigma_0)^3 
                   +\frac{\lambda}{12}(g_{\sigma}\sigma_0)^4\right]   \\
P &=& \frac{1}{3}\langle T_{ii} \rangle  
 =  \frac{1}{3}\times2\sum_i\int f_{p_i}\frac{p_i^2}{(p_i^2 + M^{*2})^{1/2}}\frac{d^3p_i}{(2\pi\hbar)^3}+\frac{g_{\omega}^2}{2m_{\omega}^2}n^2 \nonumber \\
&+& \frac{g_{\rho}^2}{8m_{\rho}^2}(n_p-n_n)^2 -\frac{1}{2}\left[m_{\sigma}^2 \sigma_0^2 + \frac{\kappa}{3}(g_{\sigma}\sigma_0)^3 
  +\frac{\lambda}{12}(g_{\sigma}\sigma_0)^4\right].
\ea
The minimization of the grand potential $\Omega = -PV$ with respect to $\sigma_0$ (equivalent to
$\partial \varepsilon/\partial \sigma_0 = 0$ at $T=0$ and to $\partial P/\partial \sigma_0=0$ at finite temperature)
leads to a self-consistent equation for the Dirac effective mass
\be
M^* = M -\frac{g_{\sigma}^2}{m_{\sigma}^2}\left[n_s-\frac{\kappa}{2}(M-M^*)^2-\frac{\lambda}{6}(M-M^*)^3\right].
\ee
In the present work we use the masses $M=939.0$ MeV, $m_{\sigma}=511.2$ MeV, $m_{\omega}=783.0$ MeV, $m_{\rho}=770.0$ MeV
and the couplings $g_{\sigma}=9.061$, $g_{\omega}=10.55$, $g_{\rho}=7.475$, $\kappa=9.194$ MeV, $\lambda=-3.280\times10^{-2}$.
These correspond to a cold symmetric nuclear matter equilibrium density $n_0=0.155$ fm$^{-3}$ at which the energy per particle
$E/A = -16$ MeV, the compression modulus $K_0 = 225$ MeV, and the symmetry energy $S_v = 30$ MeV.

\section*{References}

\bibliography{mdyi}

\begin{thebibliography}{10}
\expandafter\ifx\csname url\endcsname\relax
  \def\url#1{\texttt{#1}}\fi
\expandafter\ifx\csname urlprefix\endcsname\relax\def\urlprefix{URL }\fi
\expandafter\ifx\csname href\endcsname\relax
  \def\href#1#2{#2} \def\path#1{#1}\fi

\bibitem{flt}
G.~Baym, C.~Pethick, Landau Fermi-Liquid Theory, Wiley Interscience, New York,
  1991.

\bibitem{LB86}
A.~{Burrows}, J.~M. {Lattimer}, {The birth of neutron stars}, \apj 307 (1986)
  178--196.
\newblock \href {http://dx.doi.org/10.1086/164405} {\path{doi:10.1086/164405}}.

\bibitem{Burrows88}
A.~{Burrows}, {Supernova neutrinos}, \apj 334 (1988) 891--908.
\newblock \href {http://dx.doi.org/10.1086/166885} {\path{doi:10.1086/166885}}.

\bibitem{Sekiguchi11}
Y.~Sekiguchi, K.~Kiuchi, K.~Kyutoku, M.~Shibata,
  \href{http://link.aps.org/doi/10.1103/PhysRevLett.107.051102}{Gravitational
  waves and neutrino emission from the merger of binary neutron stars}, Phys.
  Rev. Lett. 107 (2011) 051102.
\newblock \href {http://dx.doi.org/10.1103/PhysRevLett.107.051102}
  {\path{doi:10.1103/PhysRevLett.107.051102}}.
\newline\urlprefix\url{http://link.aps.org/doi/10.1103/PhysRevLett.107.051102}

\bibitem{LLI}
L.~Landau, E.~M. Lifshitz, Statistical Physics, Volume 5, Part 1, 3rd Edition,
  Pergamon, New York, 1980.

\bibitem{ll9}
E.~M. Lifshitz, L.~P. Pitaevskii, Statistical Physics Part 2, Butterworth
  Heinemann, Oxford, 1980.

\bibitem{BaymChin76}
G.~Baym, S.~A. Chin,
  \href{http://www.sciencedirect.com/science/article/pii/0375947476905133}{Landau
  theory of relativistic fermi liquids}, Nuclear Physics A 262~(3) (1976) 527
  -- 538.
\newblock \href
  {http://dx.doi.org/http://dx.doi.org/10.1016/0375-9474(76)90513-3}
  {\path{doi:http://dx.doi.org/10.1016/0375-9474(76)90513-3}}.
\newline\urlprefix\url{http://www.sciencedirect.com/science/article/pii/0375947476905133}

\bibitem{BF81}
J.~Blaizot, B.~Friman,
  \href{http://www.sciencedirect.com/science/article/pii/0375947481900877}{On
  the nucleon effective mass in nuclear matter}, Nuclear Physics A 372~(1–2)
  (1981) 69 -- 89.
\newblock \href
  {http://dx.doi.org/http://dx.doi.org/10.1016/0375-9474(81)90087-7}
  {\path{doi:http://dx.doi.org/10.1016/0375-9474(81)90087-7}}.
\newline\urlprefix\url{http://www.sciencedirect.com/science/article/pii/0375947481900877}

\bibitem{CB93}
D.~Coffey, K.~S. Bedell,
  \href{http://link.aps.org/doi/10.1103/PhysRevLett.71.1043}{Nonanalytic
  contributions to the self-energy and the thermodynamics of two-dimensional
  fermi liquids}, Phys. Rev. Lett. 71 (1993) 1043--1046.
\newblock \href {http://dx.doi.org/10.1103/PhysRevLett.71.1043}
  {\path{doi:10.1103/PhysRevLett.71.1043}}.
\newline\urlprefix\url{http://link.aps.org/doi/10.1103/PhysRevLett.71.1043}

\bibitem{cons15}
C.~Constantinou, B.~Muccioli, M.~Prakash, J.~M. Lattimer, Phys. Rev. C, in
  press; arXiv:1504.03982v1.

\bibitem{Hama:90}
S.~Hama, B.~C. Clark, E.~D. Cooper, H.~S. Sherif, R.~L. Mercer,
  \href{http://link.aps.org/doi/10.1103/PhysRevC.41.2737}{Global dirac optical
  potentials for elastic proton scattering from heavy nuclei}, Phys. Rev. C 41
  (1990) 2737--2755.
\newblock \href {http://dx.doi.org/10.1103/PhysRevC.41.2737}
  {\path{doi:10.1103/PhysRevC.41.2737}}.
\newline\urlprefix\url{http://link.aps.org/doi/10.1103/PhysRevC.41.2737}

\bibitem{Danielewicz02}
P.~Danielewicz, R.~Lacey, W.~G. Lynch,
  \href{http://www.sciencemag.org/content/298/5598/1592.abstract}{Determination
  of the equation of state of dense matter}, Science 298~(5598) (2002)
  1592--1596.
\newblock \href
  {http://arxiv.org/abs/http://www.sciencemag.org/content/298/5598/1592.full.pdf}
  {\path{arXiv:http://www.sciencemag.org/content/298/5598/1592.full.pdf}},
  \href {http://dx.doi.org/10.1126/science.1078070}
  {\path{doi:10.1126/science.1078070}}.
\newline\urlprefix\url{http://www.sciencemag.org/content/298/5598/1592.abstract}

\bibitem{Demorest}
P.~Demorest, T.~Pennucci, S.~Ransom, M.~Roberts, J.~Hessels, A two-solar-mass
  neutron star measured using shapiro delay, Nature 467~(7319) (2010)
  1081--1083.
\newblock \href {http://dx.doi.org/10.1038/nature09466}
  {\path{doi:10.1038/nature09466}}.

\bibitem{Antoniadis}
J.~Antoniadis, P.~C.~C. Freire, N.~Wex, T.~M. Tauris, R.~S. Lynch, M.~H. van
  Kerkwijk, M.~Kramer, C.~Bassa, V.~S. Dhillon, T.~Driebe, J.~W.~T. Hessels,
  V.~M. Kaspi, V.~I. Kondratiev, N.~Langer, T.~R. Marsh, M.~A. McLaughlin,
  T.~T. Pennucci, S.~M. Ransom, I.~H. Stairs, J.~van Leeuwen, J.~P.~W.
  Verbiest, D.~G. Whelan,
  \href{http://www.sciencemag.org/content/340/6131/1233232.abstract}{A massive
  pulsar in a compact relativistic binary}, Science 340~(6131).
\newblock \href {http://dx.doi.org/10.1126/science.1233232}
  {\path{doi:10.1126/science.1233232}}.
\newline\urlprefix\url{http://www.sciencemag.org/content/340/6131/1233232.abstract}

\bibitem{Welke88}
G.~M. Welke, M.~Prakash, T.~T.~S. Kuo, S.~Das~Gupta, C.~Gale,
  \href{http://link.aps.org/doi/10.1103/PhysRevC.38.2101}{Azimuthal
  distributions in heavy ion collisions and the nuclear equation of state},
  Phys. Rev. C 38 (1988) 2101--2107.
\newblock \href {http://dx.doi.org/10.1103/PhysRevC.38.2101}
  {\path{doi:10.1103/PhysRevC.38.2101}}.
\newline\urlprefix\url{http://link.aps.org/doi/10.1103/PhysRevC.38.2101}

\bibitem{Das03}
C.~B. Das, S.~Das~Gupta, C.~Gale, B.-A. Li,
  \href{http://link.aps.org/doi/10.1103/PhysRevC.67.034611}{Momentum dependence
  of symmetry potential in asymmetric nuclear matter for transport model
  calculations}, Phys. Rev. C 67 (2003) 034611.
\newblock \href {http://dx.doi.org/10.1103/PhysRevC.67.034611}
  {\path{doi:10.1103/PhysRevC.67.034611}}.
\newline\urlprefix\url{http://link.aps.org/doi/10.1103/PhysRevC.67.034611}

\bibitem{Reinhard99}
P.-G. Reinhard, D.~J. Dean, W.~Nazarewicz, J.~Dobaczewski, J.~A. Maruhn, M.~R.
  Strayer, \href{http://link.aps.org/doi/10.1103/PhysRevC.60.014316}{Shape
  coexistence and the effective nucleon-nucleon interaction}, Phys. Rev. C 60
  (1999) 014316.
\newblock \href {http://dx.doi.org/10.1103/PhysRevC.60.014316}
  {\path{doi:10.1103/PhysRevC.60.014316}}.
\newline\urlprefix\url{http://link.aps.org/doi/10.1103/PhysRevC.60.014316}

\bibitem{Prakash88b}
M.~Prakash, T.~T.~S. Kuo, S.~Das~Gupta,
  \href{http://link.aps.org/doi/10.1103/PhysRevC.37.2253}{Momentum dependence,
  boltzmann-uehling-uhlenbeck calculations, and transverse momenta}, Phys. Rev.
  C 37 (1988) 2253--2256.
\newblock \href {http://dx.doi.org/10.1103/PhysRevC.37.2253}
  {\path{doi:10.1103/PhysRevC.37.2253}}.
\newline\urlprefix\url{http://link.aps.org/doi/10.1103/PhysRevC.37.2253}

\bibitem{Muller96}
H.~{M{\"u}ller}, B.~D. {Serot}, {Relativistic mean-field theory and the
  high-density nuclear equation of state}, Nuclear Physics A 606 (1996)
  508--537.
\newblock \href {http://arxiv.org/abs/nucl-th/9603037}
  {\path{arXiv:nucl-th/9603037}}, \href
  {http://dx.doi.org/10.1016/0375-9474(96)00187-X}
  {\path{doi:10.1016/0375-9474(96)00187-X}}.

\bibitem{jel}
S.~Johns, P.~J. Ellis, J.~Lattimer, Numerical approximation to the
  thermodynamic integrals, The Astrophysical Journal 473~(2) (1996) 1020.
\newblock \href {http://dx.doi.org/10.1086/178212} {\path{doi:10.1086/178212}}.

\bibitem{APRppr}
C.~Constantinou, B.~Muccioli, M.~Prakash, J.~M. Lattimer,
  \href{http://link.aps.org/doi/10.1103/PhysRevC.89.065802}{Thermal properties
  of supernova matter: The bulk homogeneous phase}, Phys. Rev. C 89 (2014)
  065802.
\newblock \href {http://dx.doi.org/10.1103/PhysRevC.89.065802}
  {\path{doi:10.1103/PhysRevC.89.065802}}.
\newline\urlprefix\url{http://link.aps.org/doi/10.1103/PhysRevC.89.065802}

\end{thebibliography}

\end{document}